\documentclass[fleqn,usenatbib]{mnras}

\pdfoutput=1 
             
\usepackage[T1]{fontenc}

\DeclareRobustCommand{\VAN}[3]{#2}
\let\VANthebibliography\thebibliography
\def\thebibliography{\DeclareRobustCommand{\VAN}[3]{##3}\VANthebibliography}

\usepackage{graphicx}	
\usepackage{amsmath}	
\usepackage{amssymb}	
\usepackage{scrextend}
\usepackage{lscape}

\usepackage{xcolor}
\usepackage{cleveref}

\usepackage{mathtools} 

\usepackage{newtxtext,newtxmath}

\newcommand{\fix}[1]{}

\newcommand{\thb}[1]{\left\langle{#1}\right\rangle}
\newcommand{\vecb}[1]{\boldsymbol{#1}}

\defcitealias{Balberg:2002ue}{BSI}
\defcitealias{LyndenBell:1980}{LBE}

\title[Universal gravothermal evolution for a velocity dependent cross section]{Universal gravothermal evolution of isolated self-interacting dark matter halos for velocity-dependent cross sections}

\author[N. J. Outmezguine et al.]{
Nadav Joseph Outmezguine,$^{1,2}$\thanks{E-mail: \href{mailto:njo@berkeley.edu}{NJO@Berkeley.edu}}
Kimberly K.~Boddy,$^{3}$
Sophia Gad-Nasr,$^{4}$
Manoj Kaplinghat,$^{4}$
\newauthor
and Laura Sagunski$^{5}$
\\
$^{1}$Berkeley Center for Theoretical Physics, University of California, Berkeley, CA 94720, USA\\
$^{2}$Theory Group, Lawrence Berkeley National Laboratory, Berkeley, CA 94720, USA\\
$^{3}$Theory Group, Department of Physics, The University of Texas at Austin, Austin, TX 78712\\
$^{4}$Center for Cosmology, Department of Physics and Astronomy, University of California - Irvine, Irvine, CA 92697, USA\\
$^{5}$Institute for Theoretical Physics, Goethe University, 60438 Frankfurt am Main, Germany
}

\date{Accepted XXX. Received YYY; in original form ZZZ}

\pubyear{2022}

\begin{document}
\label{firstpage}
\pagerange{\pageref{firstpage}--\pageref{lastpage}}
\maketitle

\begin{abstract}
We study the evolution of isolated self-interacting dark matter (SIDM) halos using spherically-symmetric gravothermal equations allowing for the scattering cross section to be velocity dependent. We focus our attention on the large class of models where the core is in the long mean free path regime for a substantial time. We find that the temporal evolution exhibits an approximate universality that allows velocity-dependent models to be mapped onto velocity-independent models in a well-defined way using the scattering timescale computed when the halo achieves its minimum central density. We show how this timescale depends on the halo parameters and an average cross section computed at the central velocity dispersion when the central density is minimum. The predicted collapse time is fully defined by the scattering timescale, with negligible variation due to the velocity dependence of the cross section. We derive new self-similar solutions that provide an analytic understanding of the numerical results. 
\end{abstract}



\section{Introduction}
\label{sec:intro}
Identifying universal behaviors of physical systems, such as scaling relations and self-similarity, is one of the most powerful tools in modern physics. It helps by reducing a large set of complex systems into a single essence that can be studied and explained in a unified manner. In this paper, we describe the universal evolution of self-interacting dark matter halos that is, to a good approximation, independent of the underlying thermalization process.

Thermalized self-gravitating systems are known to be unstable due to the so-called ``gravothermal catastrophe'':  The core of a gravothermal system transfers heat to its cooler and dilute outer regions, causing the core to shrink in size and heat up in a runaway process~\citep{1962spss.book.....A,Lynden-Bell:1968eqn}. The gravothermal catastrophe was originally studied in the context of globular clusters~\citep{Lightman:1978, LyndenBell:1980,1987degc.book.....S}, where the thermal ensemble was comprised of stars and the interactions were purely gravitational. \cite{Balberg:2001qg} suggested later that the gravothermal collapse of halos, due to dark matter (DM) self interactions, could be the origin of supermassive black holes. In that work, DM particles were assumed to interact as hard spheres. Both~\cite{LyndenBell:1980} and~\cite{Balberg:2002ue} (hereafter LBE and BSI respectively) identified a self-similar behavior that describes the main phase of the evolution of such gravothermal systems. In this paper, we generalize their results for a generic velocity-dependent thermalization rate, arising from a velocity-dependent self-interaction cross section. 
Surprisingly, we find that the details of the underlying microphysics can largely be removed by an appropriate scaling of time.

The practical motivation for our generalized study is to understand the cosmological evolution of isolated self-interacting DM (SIDM)  halos. SIDM was originally introduced by~\cite{Spergel_2000} to explain an apparent mismatch: halos produced in small-scale CDM-only  simulations appeared much denser than observed ones~\citep{Flores_1994, Moore_1999,deBlok_2001,Gentile:2004tb,Oh_2011,Salucci:2018hqu}. SIDM has since been shown to provide a natural account for many of the observed small-scale halo properties~\citep{Weinberg_2015,Tulin_2018}. Notably, SIDM can explain the uniformity and diversity of galactic rotation curves~\citep{Oman_2015,Kamada_2017,Ren:2018jpt}.

Given various SIDM studies across a wide range of astrophysical systems, it has become clear that the self-interaction rate must differ between systems of different scales. For example, the cross section required to address the diversity of rotation curves is larger than what is allowed by inferences of the central densities of galaxy clusters~\citep{Kaplinghat:2015aga,Elbert_2018,Sagunski_2021,Andrade:2021} and wobbling of the central cluster galaxy~\citep{Harvey:2018uwf}. A large enough SIDM interaction rate can drive dwarf-scale halos into a core-collapse phase~\citep{Elbert_2015}, yielding diverse central densities that may be constrained with the observed Milky Way satellites~\citep{Nishikawa_2020,Kahlhoefer:2019,Zavala:2019,Kaplinghat:2019svz,Correa:2021,Turner:2021,jiang:2021,Zeng:2021ldo}. 

A key observation is that the velocity dispersion of DM particles in clusters is of order $\sim 10^3\;\rm km/s$, while in galaxies, it is of order $\sim1- 100\;\rm km/s$. A velocity-dependent cross section could, therefore, permit {cross sections of order $\sim 1~\text{cm}^2/\text{g}$} for galaxy-scale halos, while maintaining consistency with cluster-scale observations, {which require cross sections $\lesssim 0.1~\text{cm}^2/\text{g}$}~\citep{Loeb:2010gj,Kaplinghat:2015aga,Sagunski_2021}. \fix{The cross section models considered in this work (based on Yukawa interaction) are able to accommodate this velocity dependence~\citep{jiang:2021}.}

The velocity dependence of the self-interaction cross section serves as a guideline for particle SIDM model building.  Velocity-dependent interactions arise  naturally in models with Yukawa or Rutherford-like long-range interactions, mediated by a light boson~\citep{Feng:2009mn,Loeb:2010gj,Tulin:2013teo}.
They could also arise via resonant scattering, without introducing any light particles~\citep{Chu:2018fzy,Chu:2019awd,tsai2020resonant}. It is interesting to note that while a light mediator exchange induces an anisotropic velocity-dependent cross section, resonant {$s$-wave} scattering is isotropic. 
From a particle physics standpoint, non-relativistic velocity-dependent interactions could be a hint for the existence of a rich dark sector, comprised of more than one state~\citep{Tulin_2013,Boddy:2014,Kaplinghat:2014,Colquhoun:2020adl}.

For the models discussed above, the resulting SIDM interaction rate is sensitive to the SIDM particle velocity, relative to a velocity scale which is fixed once the particle physics parameters are specified. The interaction rate varies from halo to halo because the underlying velocity distributions are different. Therefore, a study of the cosmological evolution of SIDM halos should  incorporate a velocity-dependent interaction rate.  

In this paper, we investigate the gravothermal evolution of isolated SIDM halos in the presence of a velocity-dependent self interaction. We consider halos that are initially sufficiently dilute, such that the SIDM particle mean-free-path (MFP) is larger than the halo size, quantified by the Jeans length $H = {v}/\sqrt{{4\pi G\rho}}$. We refer to this regime as the long MFP (LMFP) regime. If the MFP is much smaller than $H$, the halo is in the short MFP (SMFP) regime. We find that most SIDM halos that start evolving in the LMFP regime spend most of their lifetime in the LMFP regime, entering the SMFP regime only at the very last moments prior to core collapse. The details of SMFP SIDM halos are not discussed here and deferred to future work. We find that while in the LMFP regime, the evolution of SIDM halos is approximately universal. To a large extent, we find that all particle physics models can be mapped onto a velocity-independent model, once an appropriate timescale is identified.

In \S\ref{sec:gravothermal_evolution_of_isolated_halos}, we present the gravothermal evolution formalism for a  generalized treatment of the effect of velocity dependence, including a discussion of the initial conditions, particle physics models, and our numerical calculations. Our numerical results, showcasing the approximate universal behavior of isolated SIDM halos, are presented and discussed in \S\ref{sec:Core_Evolution}. In \S\ref{sec:self_similar_long_mean_free_path}, we present an analytical account of some of our findings, based on a self-similar solution to the LMFP equations that we derive for velocity-dependent self interactions. Possible implications, along with caveats and limitations of our analysis are presented as part of our conclusions in \S\ref{sec:conclusions}. This paper is supplemented by a derivation from first principles of the SMFP heat conduction term in~Appendix~\ref{sec:drivation_of_the_short_mean_free_path_heat_conduction}. We derive the short-distance expansion of a self-similar solution in Appendix~\ref{sec:short_dist_expansion} and discuss the validity of the self-similar solution in Appendix~\ref{sec:validity_of_the_lmfp_self_similar_solution}. Appendix~\ref{sec:figures_table} contains supplementary figures and a table summarizing the the parameters used in our numerical calculations.

\section{Gravothermal Evolution of Isolated Halos} 
\label{sec:gravothermal_evolution_of_isolated_halos}
In the presence of an efficient heat conduction mechanism, {assuming DM is well described as a monatomic ideal gas}, the dynamics of a spherical {non-rotating} halo  can be described by the following set of gravothermal evolution equations~\citep{1987degc.book.....S,LyndenBell:1980,Balberg:2002ue,Nishikawa_2020,Essig:2018pzq}:
\begin{align}\label{eq:gravothermal}
	\frac{\partial M}{\partial r}&=4\pi r^2\rho\;\;, \;\;\frac{\partial(\rho v^2)}{\partial r}=-\frac{GM\rho}{r^2}\;\;,\;\;\frac{L}{4\pi r^2}=-\kappa\frac{\partial T}{\partial r}\;\;,\nonumber\\
	\frac{\partial L}{\partial r}&=-4\pi r^2\rho v^2\left(\frac{\partial}{\partial t}\right)_M \mathrm{log}\left(\frac{v^3}{\rho}\right),
\end{align}
where $\rho$, $M$, $L$, and $v$ are the halo density, mass enclosed within a radius $r$, luminosity, and 1D velocity dispersion, respectively. The derivative with respect to time $t$ is performed holding $M$ fixed. The 1D velocity dispersion is related to the DM temperature via $T_{\rm dm}=v^2m_{\rm dm}$, where $m_{\rm dm}$ is the DM particle mass.

The particle nature of an SIDM model is manifested by the heat conductivity term, $\kappa$, through the relation between the luminosity and the temperature gradient (also known as Fourier's law). Prior studies have used expressions for $\kappa$ that are valid for constant SIDM cross sections.
In this section, we generalize previous results to incorporate velocity-dependent interactions and discuss our numerical methods to solve the gravothermal equations.

\subsection{Velocity Dependent Heat Conductivity}
The heat conductivity is generally not calculable from first principles; however, it is possible to estimate $\kappa$ in the limit where the MFP of DM particles is either much shorter (the SMFP regime) or much longer (the LMFP regime) than the ``size'' of the halo.
In order to cover the intermediate regime, we adopt the interpolation scheme of~\cite{Balberg:2002ue} and define the full heat conductivity through
\begin{equation}
	\frac{1}{\kappa}=\frac{1}{\kappa_{\rm LMFP}}+\frac{1}{\kappa_{\rm SMFP}}.
\end{equation}

To account for velocity-dependent interactions,
we consider a differential cross section that approaches a constant at velocities much smaller than some velocity scale $w$:
\begin{equation}\label{eq:dsig_dO}
	\lim_{v_{\rm rel}\to0}\frac{d\sigma(v_{\rm rel})}{d\Omega}=\lim_{w\to\infty}\frac{d\sigma(v_{\rm rel})}{d\Omega}=\frac{\sigma_0}{4\pi},
\end{equation}
where ${v_{\rm rel}=|\vecb{v}_1-\vecb{v}_2|}$ is the relative velocity between the incoming particles and $\sigma_0$ is a constant prefactor.
This behavior is applicable for both anisotropic (e.g., Yukawa scattering) and isotropic (e.g., s-channel resonant scattering) SIDM models.

\paragraph*{Short Mean Free Path:} 
In the SMFP regime, the local thermal properties of the SIDM fluid are agnostic to the size of the system. In this limit, the thermal conductivity can be calculated perturbatively using the Chapman-Enskog expansion~\citep{pitaevskii2012physical,chapman1990mathematical}.
The result for the case of hard-sphere scattering is well known in the literature~\citep{Koda:2011yb} and is given by $\kappa=3bv/2\sigma_0$, where $b=25\sqrt{\pi}/32$.

We present a derivation of the SMFP conductivity for a general self-interaction cross section in Appendix~\ref{sec:drivation_of_the_short_mean_free_path_heat_conduction}.  Here, we provide a more intuitive explanation of the form we adopt for $\kappa_{\rm SMFP}$. From Problem 1 of \S 10 Chapter 1 of~\cite{pitaevskii2012physical}, the heat conductivity can be expressed as
\begin{equation}\label{eq:kappa_smfp_sacle}
	\kappa_{\rm SMFP}=F(v)\thb{\sigma_{\rm visc}v_{\rm rel}^5}^{-1},
\end{equation}
{where $\sigma_{\rm visc}=\int d\sigma \sin^2\theta$ is the viscosity cross section ($\theta$ is the scattering deflection angle) and  the quantity $F(v)$ is a function of the 1D velocity dispersion and does not depend on the scattering process considered. Above $\thb{\cdot}$ denotes thermal averaging of both $\vecb{v}_{1}$ and $\vecb{v}_{2}$ over a 3D Maxwell-Boltzmann distribution with a 1D dispersion given by $v$, namely
\begin{equation}
	\thb{G(v_{\rm rel})}=\int_0^\infty \frac{dv_{\rm rel}}{\sqrt{4\pi} v^3}v_{\rm rel}^2e^{-(v_{\rm rel}/2v)^2}G(v_{\rm rel})
\end{equation} }

By construction, the differential cross section in Eq.~\eqref{eq:dsig_dO} approaches that of a hard sphere interaction for $w\to \infty$. In this limit, therefore, we require that $\kappa$ approaches its well-known form $\kappa=3bv/2\sigma_0$. We write the velocity-dependent SMFP conductivity as
\begin{equation}\label{eq:k_smfp_K_n}
	\kappa_{\rm SMFP}=\frac{3}{2}\frac{bv}{\sigma_0}\frac{1}{K_5}\;\;,\;\;K_p=\frac{\thb{\sigma_{\rm visc}v_{\rm rel}^p}}{\lim_{w\to\infty}\thb{\sigma_{\rm visc}v_{\rm rel}^p}},
\end{equation}
where we have chosen the form of $K_p$ to reflect the dependence on the cross section shown in Eq.~\eqref{eq:kappa_smfp_sacle}.

\paragraph*{Long Mean Free Path:} 
In standard kinetic theory, the LMFP regime is defined with respect to the size of the box containing the gas. As such, the resulting heat conductivity is sensitive to both the size of this box and the MFP. In a halo, however, the ``size of the box'' is not well defined, making it difficult to define the conductivity.

The procedure in the literature for computing $\kappa$ in the LMFP regime is based on dimensional analysis and matching to N-body simulations. The resulting conductivity is written as $\kappa=3 aC \sigma_0v n^2  H^2/2$, where $a=\thb{v_{\rm rel}}/v=4/\sqrt{\pi}$, $H=v/\sqrt{4\pi G\rho}$ is the Jeans scale, and $C$ is an unknown number of order unity that is calibrated to simulations~\citep{Balberg:2002ue,Koda:2011yb,1987degc.book.....S,Essig:2018pzq,LyndenBell:1980}. There have been a few attempts to determine the value of $C$~\citep{Koda:2011yb,Essig:2018pzq}, which have all relied on comparing to simulations that have implemented hard-sphere interactions.

In this manuscript, we use $C\simeq0.6$, which is shown in~\cite{Essig:2018pzq} to give better agreement with the simulations of~\cite{Koda:2011yb} in the later stages of the halo evolution, pending the collapse. We assess this choice in \S\ref{sec:Core_Evolution}. Since the halo spends the bulk of its evolution in the LMFP regime, the self-similarity and universality results presented in this manuscript can simply be re-scaled to other values of $C$. Determining the velocity dependence of $C$ is beyond the scope of this paper. 

To incorporate velocity dependence into the LMFP conductivity, we apply a similar logic to the one used for the SMFP regime. The dimensional analysis used to derive the hard-sphere result is based on the amount of energy transferred in a single SIDM collision. As shown, for example, by~\cite{Colquhoun:2020adl}, the energy transferred in such collisions is proportional to $\thb{\sigma_{\rm visc} v^3}$. Requiring once more that the heat conductivity reproduce the known result in the hard-sphere limit ($w\to\infty$), we adopt the following form for the heat conductivity:
\begin{equation}\label{eq:kappa_LMFP}
	\kappa_{\rm LMFP}=\frac{3aC}{8\pi G}\frac{\sigma_0}{m_{\rm dm}^2}\rho v^3K_3,
\end{equation}
where $K_3$ is defined in Eq.~\eqref{eq:k_smfp_K_n} for $p=3$.

It is important to note that a concrete derivation of this conductivity equation in the LMFP regime is lacking. The correct averaging over the scattering angle and relative velocity can only be determined by comparing to N-body simulations. The angular averaging resulting in the viscosity cross section is similar to the $1-|\cos\theta|$ averaging advocated by \citet{Robertson:2016qef}. However, the averaging over relative velocity in the conductivity equation has not been investigated in detail using N-body simulations. Our key conclusions are agnostic to this issue in the sense that they allow for another average to be easily used in place of the assumed $\thb{\sigma_{\rm visc} v^3}$.

\subsection{Numerical Methods} 
\label{sub:numerical_methods}
The gravothermal equations can be solved uniquely once an appropriate initial halo profile is specified. We assume dark matter self-interactions are sufficiently suppressed during structure formation, such that the halo initially evolves as a CDM halo. As a result, we assume that sufficiently early in their evolution, halos are well-described by an NFW distribution~\citep{Navarro:1997}, with a scale radius $r_s$ and scale density $\rho_s$. In this work, we analyze a variety of initially-NFW halos and particle physics parameters, as summarized in Table~\ref{table:run_params} in Appendix~\ref{sec:figures_table}. These sets of parameters yield sufficiently dilute initial halos, such that their evolution begins in the LMFP regime.

Following the method described in~\cite{Pollack:2014rja} and \cite{Nishikawa_2020}, we numerically solve the gravothermal equations in Eq.~\eqref{eq:gravothermal}, expressed in the dimensionless form
\begin{align}\label{eq:gravothermal_nondim}
	\frac{\partial \tilde{M}}{\partial \tilde{r}}&=\tilde{r}^2\tilde{\rho}\;\;, \;\;\frac{\partial(\tilde{\rho} \tilde{v}^2)}{\partial \tilde{r}}=-\frac{\tilde{M}\tilde{\rho}}{\tilde{r}^2}\;\;,\;\;{\tilde{L}}=-\tilde{r}^2\tilde{\kappa}\frac{\partial \tilde{v}^2}{\partial \tilde{r}}\;\;,\nonumber\\
	\frac{\partial \tilde{L}}{\partial \tilde{r}}&=-\tilde{r}^2\tilde{\rho} \tilde{v}^2\left(\frac{\partial}{\partial \tilde{t}}\right)_{\tilde{M}} \mathrm{log}\left(\frac{\tilde{v}^3}{\tilde{\rho}}\right),
\end{align}
where the physical quantity $x=\{\rho,v,M,L,r,t,\kappa\}$ is normalized by an appropriate scale $x_N$ to produce the dimensionless quantity $\tilde{x}\equiv x/x_N$. For consistency with Eq.~\eqref{eq:gravothermal_nondim}, the physical scales obey the relations
\begin{align}\label{eq:N_scales}
	&\rho_N=\frac{M_N}{4\pi r_N^3}\;\;,\;\;v_N^2=\frac{GM_N}{r_{N}}\;\;,\;\;L_N=\frac{M_Nv_N^2}{t_N}\;\;,\nonumber\\
	&\kappa_N=\frac{M_N}{4\pi m_{\rm dm}r_Nt_N}.
\end{align}
We find the following choice of time scale convenient:
\begin{equation}\label{eq:t_N}
	t_N= \frac{2}{3aC}\left(\rho_{N}\frac{\sigma_0}{m_{\rm dm}}v_{N} K_3\left(\frac{v_{N}}{w}\right)\right)^{-1}.
\end{equation}
Equations~\eqref{eq:N_scales} and~\eqref{eq:t_N} provide five relations between seven physical scales, meaning that once we set two scales, the remaining five automatically follow. In practice, we use the NFW scale density $\rho_s$ and scale radius $r_s$; however, other normalization choices are equally valid, and we exercise this flexibility in \S\ref{sec:self_similar_long_mean_free_path} to present our results in a more intuitive manner.

\begin{figure*}
\centering
\includegraphics[width=0.49\textwidth]{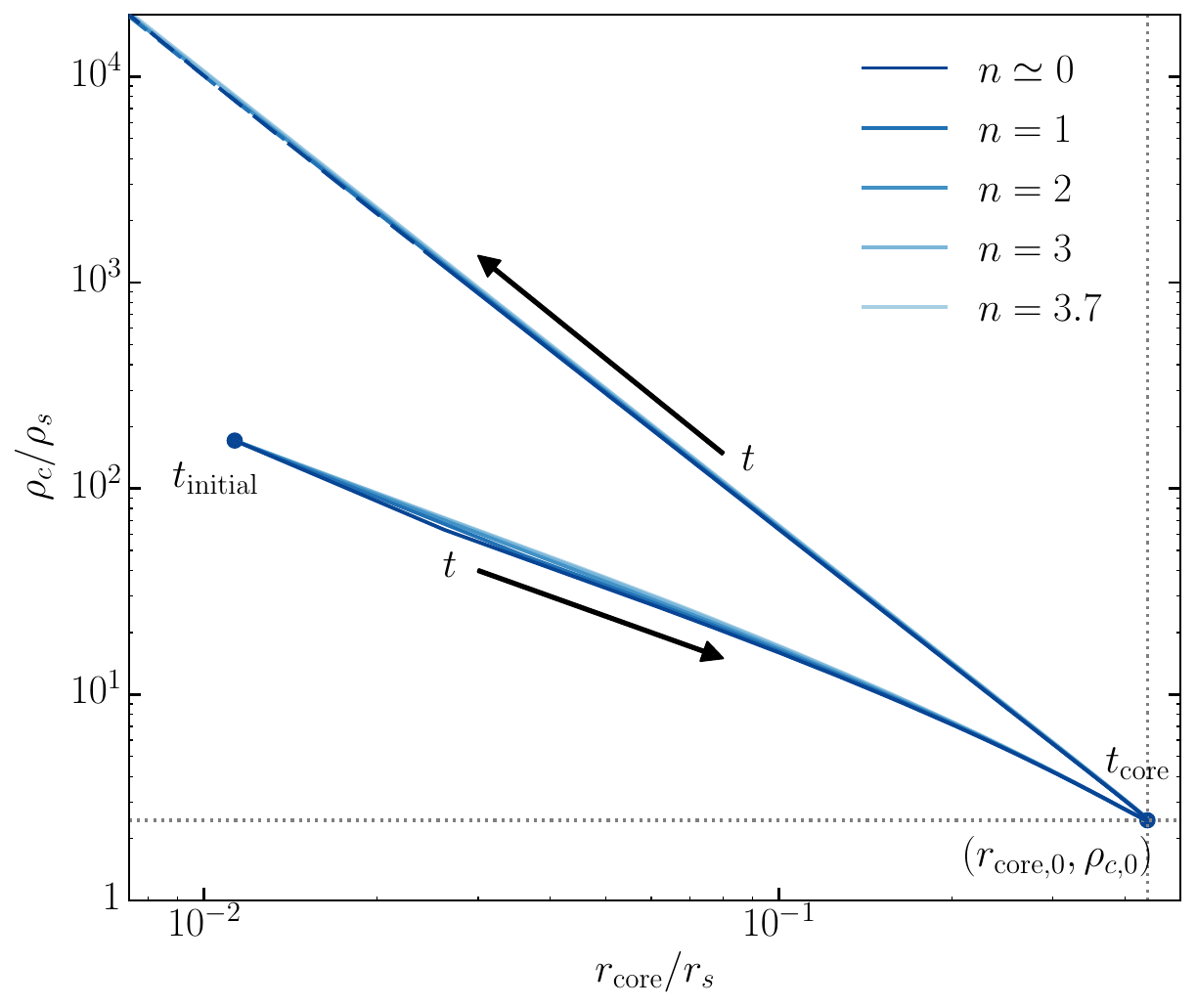}\;\;\;\;\includegraphics[width=0.49\textwidth]{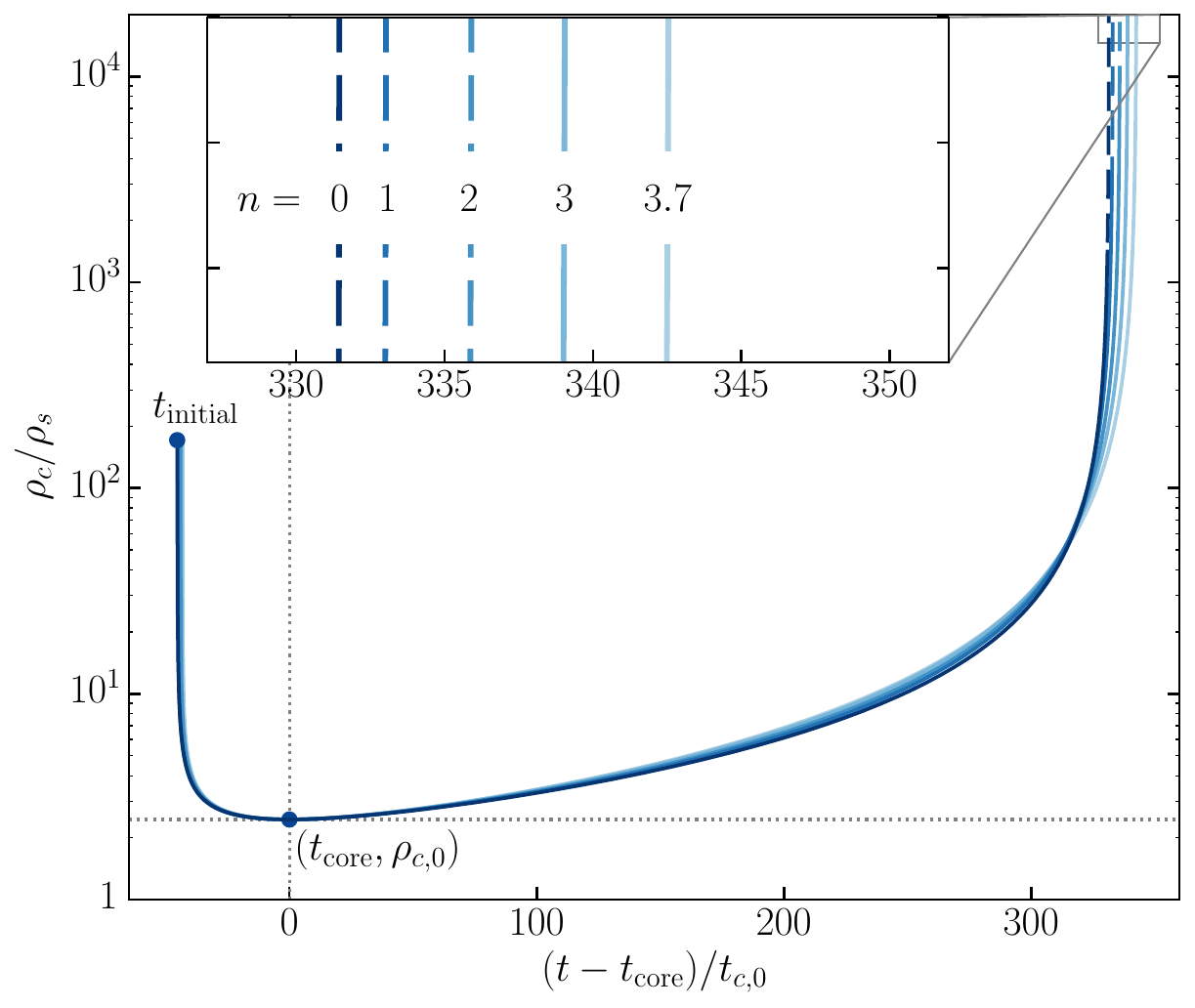}
\caption{
{\it (Left)} The halo central density, normalized by the NFW parameter $\rho_s$, as a function of the core radius, normalized by the NFW parameter $r_s$. Two instances in time are indicated: $t_{\mathrm{initial}}$, the time corresponding to the initial time step in our gravothermal evolution code, and $t_{\mathrm{core}}$, the time at which the core reaches a maximal size and minimal density.  Arrows indicate the direction of the flow of time, showing that as the evolution progresses, the central density rapidly increases after reaching its minimum, while the core radius decreases after reaching its maximum. 
{\it (Right)} Central density, normalized by $\rho_s$, as a function of shifted time, normalized by $t_{c,0}$ given in Eq.~\eqref{eq:t_N}, for the same models shown in the left panel. The inset is expanded to show the difference in the collapse times for each n of our main models. In both panels, we show results for our main models with $n=0,1,2,3,3.7$ [cf. Eq.~\eqref{eq:def_n}] for $\sigma_{c,0}/m_\mathrm{dm}\simeq 5 cm^2/g$ [cf. Eq.~\eqref{eq:sig_c0}]. The model parameters used are listed in the top five rows of Table~\ref{table:run_params}, and lines with lighter shades of blue correspond to larger values of $n$. A given line becomes dashed when $\kappa_{\mathrm{SMFP}}<\kappa_{\mathrm{LMFP}}$, meaning the core has entered the SMFP regime. The dotted gray lines cross at $t_{\rm core}$, and highlight the maximal core size, $r_{\rm core,0}$  and minimal central density $\rho_{c,0}$.  
}
\label{fig:rhoc_rc_vs_t_plots}
\end{figure*}

The dimensionless form of the conductivity is given by
\begin{equation}\label{eq:kappa_tild}
	\tilde{\kappa}=\tilde{\rho}\tilde{v}^3\tilde{K}_3\left[1+\hat{\sigma}^2\tilde{\rho}\tilde{v}^2\tilde{K}_3\tilde{K}_5\right]^{-1}\;\;,\;\;\tilde{K}_{p}=\frac{K_{p}(\tilde{v} / \hat{w})}{K_{p}(1/ \hat{w})},
\end{equation}
where 
\begin{equation}\label{eq:def_hat_sig}
	\hat{\sigma}^2=\frac{aC}{b}K_3\left(\frac{1}{\hat{w}}\right)K_5\left(\frac{1}{\hat{w}}\right)\left(\frac{M_{N}}{4\pi r^2_{N}}\frac{\sigma_0}{m_{\rm dm}}\right)^2\;\;,\;\;\hat{w}=\frac{w}{v_{N}}.
\end{equation}
Thus, the gravothermal equations are fully specified by two dimensionless parameters: $\hat{\sigma}$ and $\hat{w}$. Moreover, by choosing to normalize using the NFW parameters, the initial halo profile does not depend on our choice of parameters and is given by $\tilde{\rho}_{\rm initial}(\tilde{r})=\tilde{r}^{-1}(1+\tilde{r})^{-2}$. 

For both practical and presentation reasons, we find it useful and  intuitive to describe the cross section  according to the slope of its velocity dependence, $n$, instead of in terms of $\hat{w}$. We define $n$ such that in the vicinity of $v_N$, the cross section scales as $\thb{\sigma}\sim(v_N/v)^n$. More precisely, we define $n$ through the LMFP heat conductivity via
\begin{equation}\label{eq:def_n}
	n=3-\left.\frac{d\log \kappa_{\rm LMFP}}{d\log v}\right|_{v=v_N}.
\end{equation}

\paragraph*{Yukawa Scattering:} 
\label{par:yukawa_scattering}
Having obtained generic expressions for the conductivity in the LMFP and SMFP regimes, we now focus on the particular case of a Yukawa interaction between SIDM particles to perform the numerical calculations in this paper. Under the Born approximation, the self-interaction differential cross section can be expressed as\footnote{We note that, while commonly used in the literature, the expression given in Eq.~\eqref{eq:dsig_dOmega} is invalid for identical particles. This is most easily seen by the fact that the equation is not symmetric under taking $\theta\to\pi-\theta$. We are not concerned with this issue, as it does not effect any of our conclusions, which are qualitative in nature. Furthermore, Eq.~\eqref{eq:dsig_dOmega} holds for any spin of the SIDM particle, whereas the differential cross section for identical particles changes based on the spin. We defer the study of those details to a future work.}
\begin{equation}\label{eq:dsig_dOmega}
	\frac{d\sigma}{d\Omega}=\frac{\sigma_0}{4\pi}\left(1+\frac{v_{\rm rel}^2}{w^2}\sin^2\frac{\theta}{2}\right)^{-2}.
\end{equation}
For this cross section, $K_p$ is a function of the single variable $x \equiv v/w$, which highlights the property that $K_p$ becomes velocity-independent in the limit $w\to\infty$. The asymptotic behavior of $K_p$ is given by
\begin{equation}
	{K_p(x)}=
	\begin{cases}
	1&x\ll1\\
	\frac{3}{2(p^2-1)x^4}\left(\log(4x^2)+\psi\left(\frac{p-1}{2}\right)-2\right)&x\gg1
	\end{cases}\;\;,
\end{equation}
where $\psi$ is the Di-Gamma function. The limit $x\ll1$ ($w\gg v$) recovers the known hard-sphere scattering result by design. In the $x\gg 1$ limit, the scaling $K_p\sim \log(x)/x^4$ is reminiscent of the infrared 
divergence of Rutherford-like scattering. \fix{A plot showing the resulting $n$ [c.f. Eq.~\eqref{eq:def_n}] for the Yukawa cross section is shown in Fig.~\ref{fig:c1_comp}}

\paragraph*{Numerical Convergence Tests:} 
\label{par:numerical_convergence_tests_}
The numerical solution to the gravothermal equations (Eq.~\ref{eq:gravothermal_nondim}) is obtained by discretizing the equations over a grid of fixed mass shells as described in \cite{Pollack:2014rja,Nishikawa_2020}. 
\fix{
	To account numerically for the heat conduction, we need to take small time steps such that the density can be treated as approximately constant. To achieve this, our time step were chosen according the conditions described in appendix A of \cite{Nishikawa_2020}, with setting $\epsilon_t=10^{-4}$.
}
Our solutions are obtained for 400 logarithmically-spaced radial shells {between $\tilde{r}=10^{-2}$ and $10^2$}, and the mass within each shell is held constant throughout the evolution. We have also run models $1$ and $11$ with 800 shells to check for convergence. In going from 400 to 800 shells, we only see small differences in the results. For example, the minimum density changes by $0.016$\%, the dispersion at minimum density by $0.023$\% and the collapse time by $0.1$\%. This level of convergence is sufficient for the purpose of elucidating the approximate universality in the evolution of self-interacting dark matter halos. 

\section{Universal Gravothermal  Evolution}
\label{sec:Core_Evolution}
The specific case of hard spheres, or point-like interactions, is obtained by taking the $w\to\infty$ (or equivalently $K_p\to 1$) limit of Eqs.~\eqref{eq:gravothermal_nondim},~\eqref{eq:t_N} and~\eqref{eq:def_hat_sig}. As noted in~\citetalias{Balberg:2002ue}, the full set of gravothermal equations is sensitive to a single parameter $\hat{\sigma}$, which controls when the long-to-short mean free path transition occurs. {As $\hat{\sigma}\to0$ in Eq.~\eqref{eq:kappa_tild}, the conductivity approaches the LMFP conductivity ($\kappa \to \kappa_\text{LMFP}$), which dominates over the SMFP conductivity ($\kappa_\text{LMFP} \ll \kappa_\text{SMFP}$). Consequently, there are no free parameters at all in this LMFP phase.}
Therefore, for hard sphere interactions, the LMFP evolution of halos is trivially universal~\citepalias{Balberg:2002ue}.

As discussed in \S\ref{sec:gravothermal_evolution_of_isolated_halos}, a velocity-dependent cross section introduces a new velocity scale $\hat{w}$ (or, equivalently, $n$) to the gravothermal evolution equation, seemingly breaking universal LMFP evolution of SIDM halos. 
\begin{addmargin}[1.3em]{1.3em}
\textit{Surprisingly, we find that the evolution of SIDM halos remains largely universal when a velocity dependence is included. Moreover,  most of the important SIDM halo properties can be derived directly from the initial NFW halo parameters. At an instance in time we refer to as the ``maximal core'' stage, all halos reach the same maximal core size $r_{\rm core,0}\simeq0.45r_s$, the same minimal normalized density $\rho_{c,0}\simeq2.4\rho_s$ and the same normalized velocity dispersion $v_{c,0}\simeq0.64V_{\rm max}$.}
\end{addmargin}

\begin{figure*}
\centering
\includegraphics[width=0.49\textwidth]{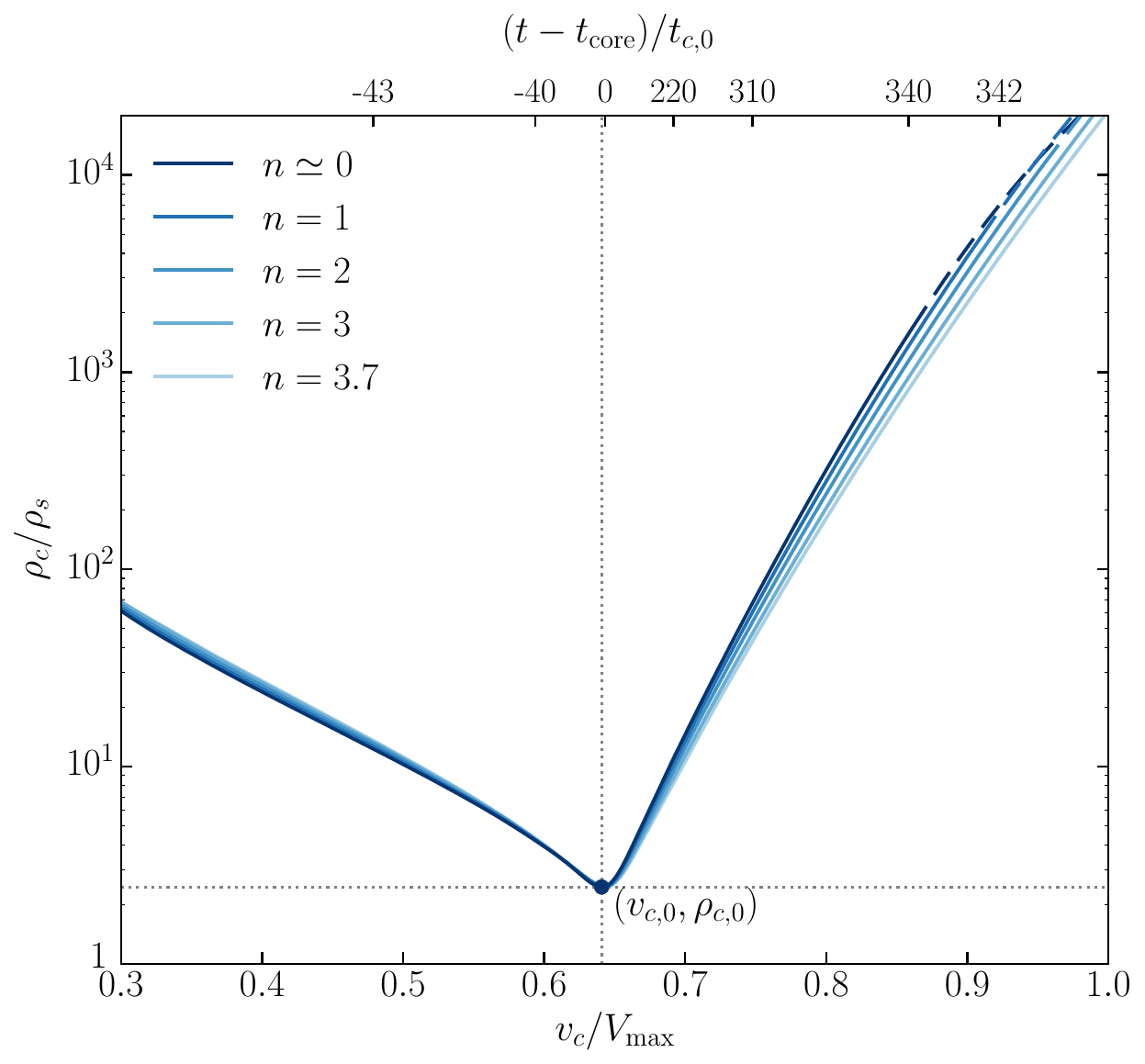}\;\;\;\;\includegraphics[width=0.49\textwidth]{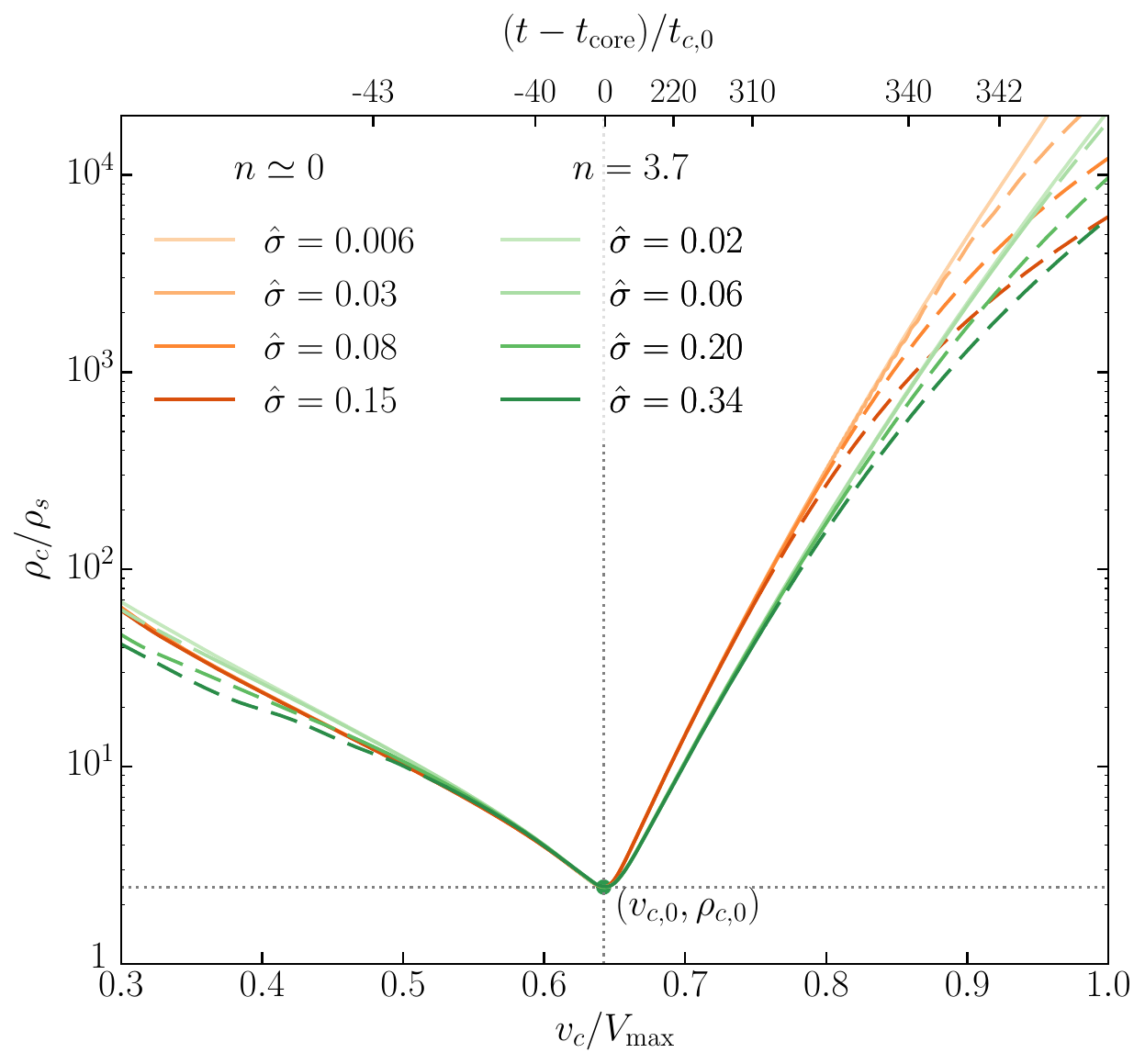}
\caption{
Central halo density, normalized by the NFW scale density $\rho_s$, as a function of the central 1D velocity dispersion, normalized by maximal NFW rotational velocity $V_{\mathrm{max}}$. In the {\it (left)} panel,  we hold $\sigma_{c,0}/m_\mathrm{dm}\simeq 5 cm^2/g$ fixed [cf. Eq.~\eqref{eq:sig_c0}] for different choices of $n$ [cf. Eq.~\eqref{eq:def_n}], for the same models shown in Fig.~\ref{fig:rhoc_rc_vs_t_plots}. In the {\it (right)} panel, we fix $n\simeq0$ and $n=3.7$, while varying over $\hat{\sigma}$. The shifted dimensionless time is shown across the top horizontal axis, demonstrating that the bulk of the evolution is spent right around core formation (we use Run \#5 of Table~\ref{table:run_params} to determine the time axis). The dotted gray lines intersect at the point at which the central density of the halo reaches its minimum value at $\rho_{c,0}$ with a corresponding 1D velocity dispersion $v_{c,0}$. Dashed lines indicate that $\kappa_{\mathrm{SMFP}}<\kappa_{\mathrm{LMFP}}$, meaning the core has entered the SMFP regime.
See Table~\ref{table:run_params} for details of each analysis run.
}
\label{fig:rhoc_vc_plots}
\end{figure*}

This conclusion is evident in  Figs.~\ref{fig:rhoc_rc_vs_t_plots} and~\ref{fig:rhoc_vc_plots}, which show the results for the various analysis runs listed in Table~\ref{table:run_params}. In the left panel of Fig.~\ref{fig:rhoc_rc_vs_t_plots}, we show the evolution of the central  density $\rho_c$ against the core size $r_{\rm core}$, defined through $\rho(r_{\rm core})=\rho_c/2$. Both are normalized by the NFW scale parameters, and the central density is defined as the density in the innermost shell resolved by our code. In  Fig.~\ref{fig:rhoc_vc_plots}, we show the evolution of the normalized central density versus the central 1D velocity dispersion, normalized by the NFW maximum rotational velocity $V_{\rm max}\simeq1.65r_s\sqrt{G\rho_s}$. It is clear that all halos evolve in a very similar manner. Most pronounced is the point of maximal core, where all lines intersect at the point where the density, 1D dispersion, and core size are given by $\rho_{c,0}\simeq2.4\rho_s,\;v_{c,0}\simeq0.64V_{\rm max},\;r_{{\rm core},0}\simeq0.45r_s$, respectively. We denote the instant in time at which a halo reaches a maximal core by $t_{\rm core}$.

As discussed in \S\ref{sec:gravothermal_evolution_of_isolated_halos}, two physical scales and two dimensionless parameters fully specify the gravothermal equations. The universality of the maximal core stage motivates us to set $v_N=v_{c,0}$ and $\rho_N=\rho_{c,0}$ as our normalizing physical scales. With these scales, we can use Eqs.~\eqref{eq:def_hat_sig} and~\eqref{eq:def_n} to calculate $\hat{\sigma}$ and $n$ for each run in Table~\ref{table:run_params}. These values are also provided in the legends in Figs.~\ref{fig:rhoc_rc_vs_t_plots} and~\ref{fig:rhoc_vc_plots}. The normalizing time scale in Eq.~\eqref{eq:t_N} is now given explicitly by
\begin{align}\label{eq: t_c0_units}
	t_{c,0}\simeq&1.5\;{\rm Gyr}\;\times\\
	&\times\left(\frac{0.6}{C}\right)\left(\frac{\rm cm^2/g}{\sigma_{c,0}/m_{\rm dm}}\right)\left(\frac{100 \rm km/s}{V_{\rm max}}\right)\left(\frac{10^7M_\odot\,\rm kpc^{-3}}{\rho_s}\right),\nonumber
\end{align}
where
\begin{equation}\label{eq:sig_c0}
	\sigma_{c,0}=\sigma_0 {K}_3\left(\frac{v_{c,0}}{w}\right)=\frac{3}{2}\left.\frac{\thb{\sigma_{\rm visc}v_{\rm rel}^3}}{\thb{v_{\rm rel}^3}}\right|_{v=v_{c,0}\simeq0.64V_{\rm max}}.
\end{equation}
The right panel of Fig.~\ref{fig:rhoc_rc_vs_t_plots} shows the time evolution of the central density, where time is shifted by $t_{\rm core}$ and normalized  by the timescale $t_{c,0}$ above. 

The near-universality of the halos evolution as a function of the dimensionless time $t/t_{c,0}$ suggests that different particle physics models that result in the same value of $\sigma_{c,0}/m_{\rm dm}$ should result in nearly identical cosmological halo evolution. This leads to the question: Is $\sigma_{c,0}/m_{\rm dm}$ the quantity constrained when we fit SIDM halos to rotation curves and other measures of halo mass profiles? In comparing to data, the two methods employed are fitting analytic SIDM halo mass profiles calibrated for constant cross sections to data~\citep{Kaplinghat:2015aga,Ren:2018jpt,Sagunski_2021,Andrade:2021} and directly comparing simulations to data~\citep{Elbert_2018,Harvey:2018uwf,Robertson_2019,Bondarenko:2020mpf}.
However, both methods have issues when velocity dependence is important: it is not clear how to interpret the cross section that results from fitting the analytic mass profiles to data, and directly comparing to simulations for a full range of velocity-dependent models is computationally expensive. Our result implies that it should be possible to implement both methods by using $\sigma_{c,0}$ as a constant cross section and relying on well-calibrated analytic models~\cite{Robertson_2020}. {See Appendix~\ref{sec:over_counting} for a more detailed discussion.}

\paragraph*{Phases of Gravothermal Evolution:} 
\label{par:phases_of_gravothermal_evolution_}
Figures~\ref{fig:rhoc_rc_vs_t_plots} and~\ref{fig:rhoc_vc_plots} show not only the universality, but also the phases of the gravothermal evolution.  The initially-NFW halo forms an inner core that gradually expands until it reaches its maximal core stage at time $t=t_{\rm core}\simeq 45t_{c,0}$.  After the maximal core is achieved, halos evolve self-similarly, as noted by~\citetalias{LyndenBell:1980} for Rutherford interactions and by~\citetalias{Balberg:2002ue} for velocity-independent cross sections. We provide a generalized discussion of the self-similar evolution in \S\ref{sec:self_similar_long_mean_free_path}.

As shown in Fig.~\ref{fig:rhoc_vc_plots}, the halo spends most of its evolution near the maximal core stage. The top horizontal axis of the plot in each panel shows time flowing from left to right. We stress that we create the time axes in Fig.~\ref{fig:rhoc_vc_plots} by mapping $v_c/V_{\rm max}$ to $t/t_{c,0}$, which depends slightly on the model parameters; we use parameters from Run \#5 of Table~\ref{table:run_params} for the conversion. The small spread of lines in Fig.~\ref{fig:rhoc_rc_vs_t_plots} helps in estimating the accuracy of the time axes of the panels in Fig.~\ref{fig:rhoc_vc_plots} for the other runs.

Self-similarity breaks down approximately when the core becomes sufficiently dense and enters the SMFP evolution, which we define as $\kappa_{\rm SMFP}<\kappa_{\rm LMFP}$ and represent in all our figures with dashed lines. In the right panel of Fig.~\ref{fig:rhoc_vc_plots}, we show the dependence on $\hat{\sigma}$ for our limiting cases of $n\simeq0$ and $n=3.7$. It is evident that the larger the $\hat{\sigma}$, the earlier the halo enters the SMFP regime. The spread of the dashed lines suggests that universality breaks down at the same point.  However, as indicated by the time axis of Fig.~\ref{fig:rhoc_vc_plots}, the transition occurs in the very late stages of the evolution where $t-t_{\rm core}\gtrsim 310 t_{c,0}$. The same conclusion is evident in right panel of Fig.~\ref{fig:rhoc_rc_vs_t_plots}, where lines become dashed in the very late stages of the evolution. See Figs.~\ref{fig:rhoc_rc_vs_t_app} and~\ref{fig:vls_sighat} in Appendix~\ref{sec:validity_of_the_lmfp_self_similar_solution} for further exploration of the effect of varying $\hat{\sigma}$ on the halo evolution. 

As seen in Fig.~\ref{fig:rhoc_rc_vs_t_plots}, the core of the halo collapses (i.e., $r_{\rm core}\to0$ and $\rho_c\to\infty$\footnote{\fix{Numerically, needless to say, it is impassible to resolve infinite density or vanishing cores. In practice we stop the numerical evolution when the core density change in a single time step exceeds $d\log\rho_c/d\log t=10^3$.}}) at a time $t=t_{\rm coll}\simeq 335 t_{c,0}$. This point is worth emphasizing:
\begin{addmargin}[1.3em]{1.3em}
\textit{We find that all halos collapse at practically the same dimensionless time $t_{\rm coll}/t_{c,0} \simeq 335$, where the timescale $t_{c,0}$ is determined directly by the initial NFW halo parameters and the cross section.}
\end{addmargin}

We refer the reader to Fig.~\ref{fig:v_rho_logslope} in Appendix~\ref{sec:figures_table}, where we show the time evolution of the full velocity and density profiles, alongside  the the log-slopes of the density profiles, for our main $n\simeq0$ and $n=3.7$ models.

\paragraph*{Comparison to N-Body Simulations} 
\label{par:comparison_to_n_body_simulations}
To validate our results, we compare to~\citet{Koda:2011yb}, who preformed N-body simulations for the evolution of isolated halos with SIDM interacting as hard spheres. Their simulations assumed a couple of different initial halo profiles. Relevant for us are the cases with an initial NFW halo or an initial halo given by the self-similar solution of~\citetalias{Balberg:2002ue}. The  simulations of \citet{Koda:2011yb} starting with the profile of~\citetalias{Balberg:2002ue} resulted in a core collapse happening at $t_{\rm coll}\simeq 385(a \rho_{c,0}v_{c,0}\sigma_0/m_{\rm dm})^{-1}$. Our numerical results (shown in Fig.~\ref{fig:rhoc_rc_vs_t_plots}) indicate a collapse of $n\simeq 0$ halos at $t_{\rm coll}\simeq331t_{c,0}$. For the two collapse times to agree, we must set $C\simeq0.57$. \cite{Essig:2018pzq} performed a similar comparison between their gravothermal code and the simulations of~\citet{Koda:2011yb}, finding $C\simeq0.6$.

The simulations of~\citet{Koda:2011yb} starting with an initial NFW profile result in a core density evolution very similar to the ones we obtain, as shown in the right panel of Fig.~\ref{fig:rhoc_rc_vs_t_plots}. Their simulations, however, reach a minimal density lower than our calculations by roughly $20\%$. We find $\rho_{c,0}\simeq2.4\rho_s$, while~\citet{Koda:2011yb} show $\rho_{c,0}\simeq 2\rho_s$. Rescaling our central density by $2/2.4$ and setting $C=0.61$ results in a remarkable agreement of our results and the NFW run shown in Fig.~6 of~\citet{Koda:2011yb}, throughout the halo evolution. 

We note that~\citet{Koda:2011yb} also report a $20\%$ difference between the minimal density they obtain and the minimal densities obtained in the gravothermal results of~\citetalias{Balberg:2002ue}. Instead of rescaling the central density, as we advocate here, \citet{Koda:2011yb} shifted the \citetalias{Balberg:2002ue} results in time, such that the two agree in the late phase of gravothermal evolution. This procedure resulted, however, in a significantly  different core formation in the simulation compared to the gravothermal code.

\section{Self-Similar Long Mean-Free-Path Evolution of SIDM Halos} 
\label{sec:self_similar_long_mean_free_path}

\begin{figure*}
\centering
\includegraphics[width=0.48\textwidth]{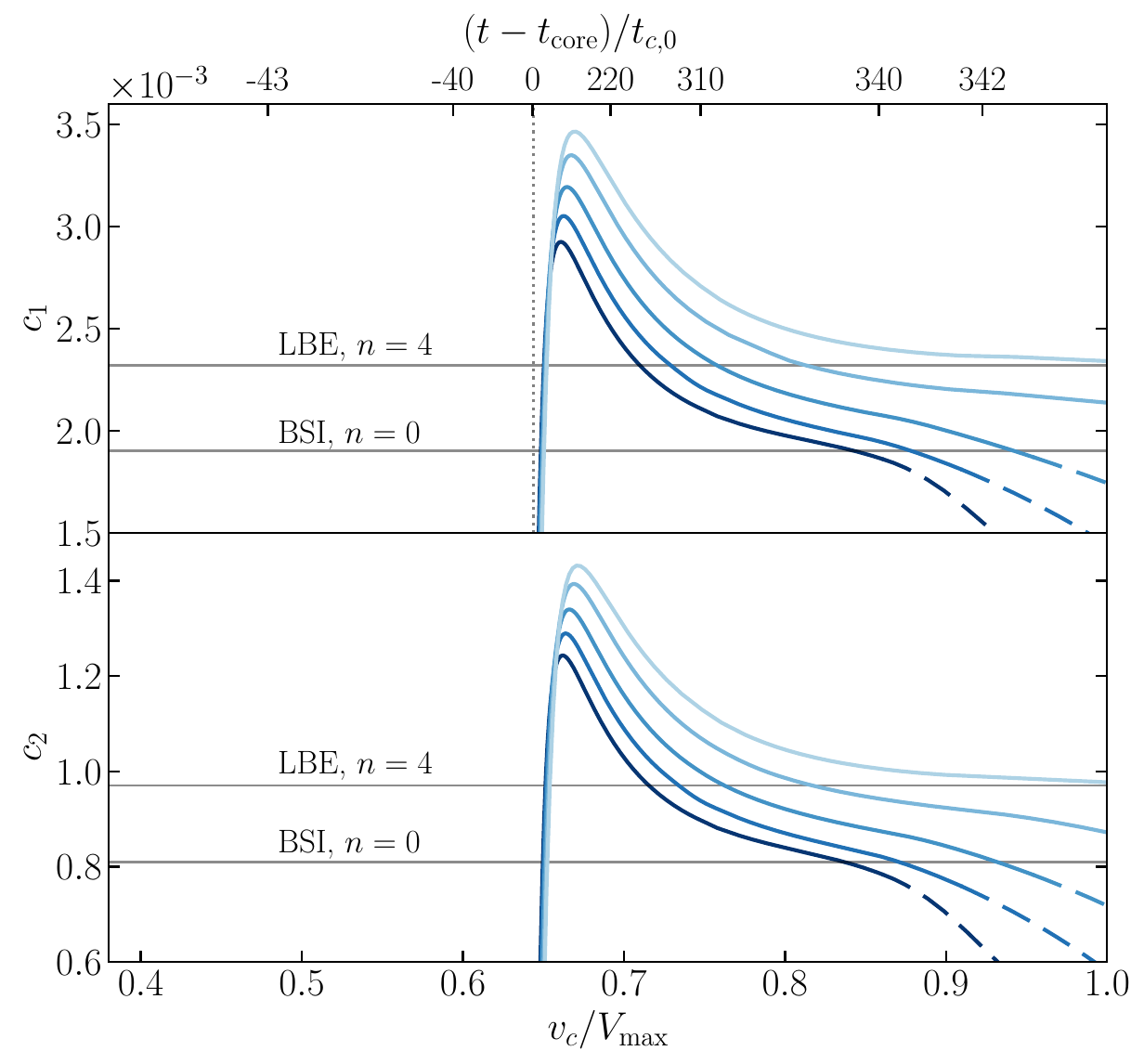}\;\;\;\;\includegraphics[width=0.49\textwidth]{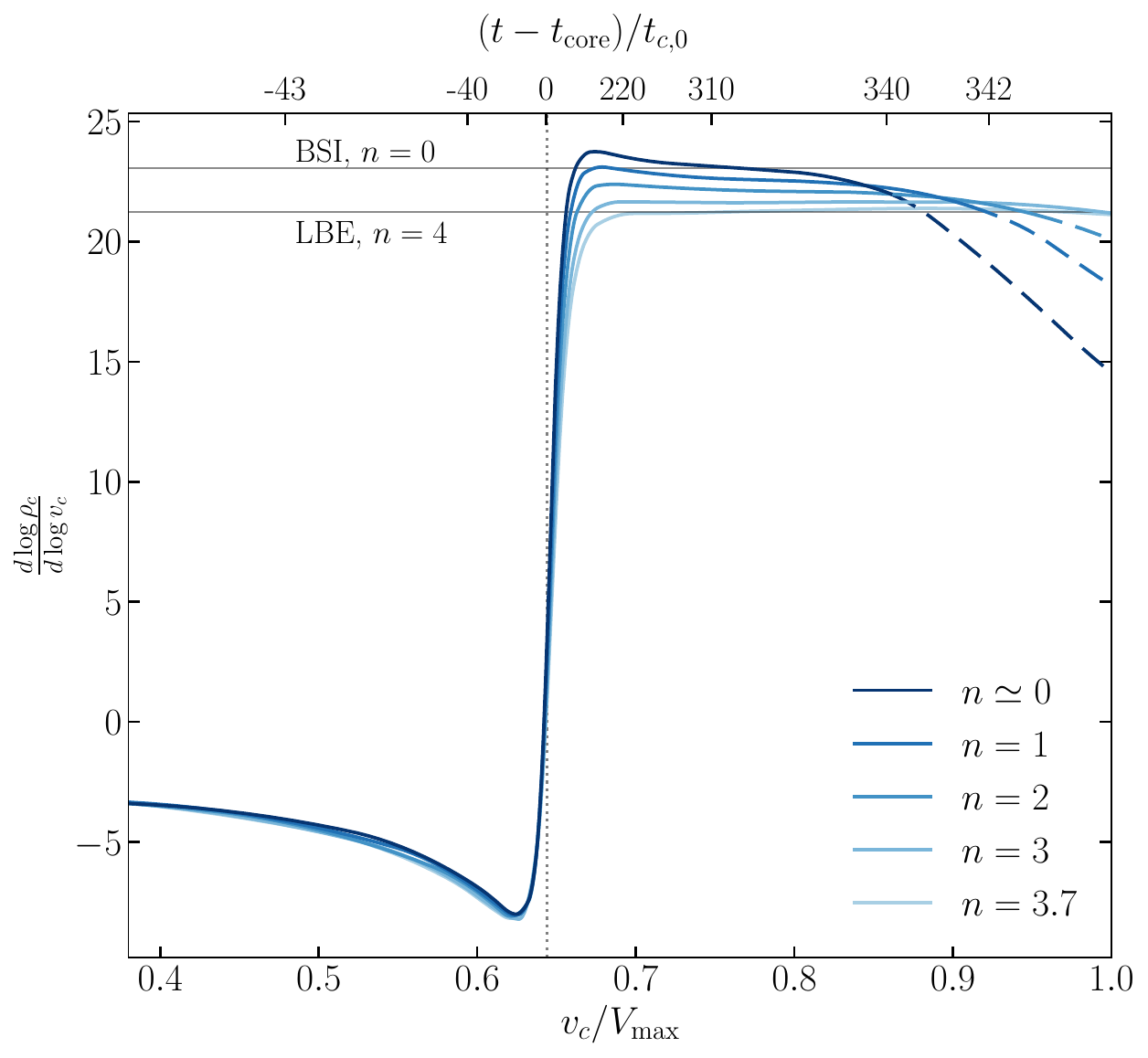}
\caption{\ 
{\it(Left)} The {separation-of-variables constants} $c_1$ and $c_2$ computed using Eq.~\eqref{eq:def_c1_c2}, plotted as a function of the central 1D velocity dispersion, normalized by maximal NFW rotational velocity $V_{\mathrm{max}}$.  Solid gray lines represent the values of $c_{1,2}$ predicted in the literature for the self-similar phase: $c_1=2.322\times10^{-3},\,c_2=9.704\times10^{-4}$ for $n=4$~\citepalias{LyndenBell:1980} and $c_1=1.903\times10^{-3},\,c_2=8.092\times10^{-4}$ for $n=0$~\citepalias{Balberg:2002ue}. {\it (Right)} Log derivative of the central density with respect to the central velocity dispersion. The solid gray lines indicate the predicted values  obtained through Eq.~\eqref{eq:alpha_logslope_relation}, using values of $\alpha$ predicted in the literature for the self-similar phase; $\alpha=2.208$ for $n=4$~\citepalias{LyndenBell:1980} and $\alpha=2.190$ for $n=0$~\citepalias{Balberg:2002ue}. In both panels, the dotted line indicates the time at which the central density reaches a minimum, $t_{\mathrm{core}}$, as well as the central velocity dispersion at that time, $v_{c,0}$. The shifted time normalized by $t_{c,0}$ [c.f. Eq.~\eqref{eq: t_c0_units}] is shown across the top horizontal axis (we use Run \#5 of Table~\ref{table:run_params} to determine the time axis). In both panels, we hold $\sigma_{c,0}/m_\mathrm{dm}\simeq 5 cm^2/g$ fixed [cf. Eq.~\eqref{eq:sig_c0}] for different choices of $n$ [cf. Eq.~\eqref{eq:def_n}], for the same models shown in Fig.~\ref{fig:rhoc_rc_vs_t_plots}. {The values of $\alpha$ for each of the runs as per Eq.~\ref{eq:alpha_logslope_relation} are: $\alpha = 2.192,2.195,2.199,2.204,2.207$ for the cases of $n=0,1,2,3,3.7$ respectively}. 
}
\label{fig:c1c2_logslope_plots}
\end{figure*}

{This section is devoted to developing a  physical intuition for our results. We present analytical tools that help explaining some of the main results presented in the previous section, including the different timescales and explanations of universal and non-universal features of the LMFP evolution of halos.} We derive a self-similar solution for the LMFP evolution, which allows us to estimate the lifetime of gravothermal halos and the region of validity of our analytical results.

For the discussion in this section, we set $t_{\mathrm{core}} = 0$, such that $t=0$ corresponds to the time of maximal core, rather than the initial time we begin gravothermal evolution in our code. Therefore, the central density and 1D velocity dispersion of the halo at $t=0$ is given by $\rho_{c,0}$ and $v_{c,0}$, respectively.

\paragraph*{Self similar evolution equation:} 
\label{sub:self_similar_evolution_equation}
Following our findings presented in the previous section, we set $v_N=v_{c,0}$, $\rho_N=\rho_{c,0}$, and $t_N = t_{c,0}$ as defined in Eq.~\eqref{eq: t_c0_units}; the remaining normalization scales follow from Eq.~\eqref{eq:N_scales}. In the LMFP regime ($\hat{\sigma}\to0$), the dimensionless conductivity of Eq.~\eqref{eq:kappa_tild} is given by $\tilde{\kappa}=\tilde{\rho}\tilde{v}^3\tilde{K}_3$. Devising a self-similar solution is impossible for a heat conductivity that is not a power law of the other variables; therefore, to make progress, we need to adopt an approximate form for $\tilde{K}_3$. Recalling the definition of $n$ in Eq.~\eqref{eq:def_n} suggests that $\tilde{\kappa}\sim\tilde{v}^{3-n}$ should be valid as long as $v/v_{c,0}$ is of order unity. As seen in Fig.~\ref{fig:rhoc_vc_plots}, $v_c/v_{c,0}$ changes only very mildly over the course of the halo evolution,  suggesting that the approximation $\tilde{\kappa}\sim\tilde{v}^{3-n}$ should be valid at the halo core throughout the evolution. This approximation breaks outside the core, but we find it captures most of the relevant halo properties. By construction, $\tilde{K}_p=1$ at $\tilde{t}=\tilde{r}=0$; therefore, we adopt the following LMFP effective conductivity when discussing the self-similar halo evolution:
\begin{equation} \label{eq:mn_eq}
	\tilde{\kappa}_{\rm eff}=\tilde{\rho}\tilde{v}^{3-n},
\end{equation}
where $n$ is defined in Eq.~\eqref{eq:def_n}. Regardless of the choice of parameters, $0<n<4$ for Yukawa interactions. We stress that our numerical results are based on the full functional form of $K_p$; the above approximation is only adopted when discussing the self-similar solution.

With the above approximated heat conduction, we follow~\citetalias{LyndenBell:1980} to seek a self-similar solution of the form
\begin{equation}\label{eq:self_similar_X}
	\tilde{x}(\tilde{r},\tilde{t})=\tilde{x}_c(\tilde{t})x_*(r_*)\;\;,\;\; r_*=\frac{\tilde{r}}{\tilde{r}_c(\tilde{t})},
\end{equation}
where $x=\{\rho,M,v,L\}$. We impose the boundary conditions
\begin{equation}\label{eq:BC_IC}
	\rho_*(0)=v_*(0)=1\;\;\text{and}\;\;L_*(0)=M_*(0)=0
\end{equation}
at $r= r_\ast = 0$. The initial condition $\tilde{x}_c(0) =1$ for central halo quantities then follows immediately from our choice of normalization scales $v_{c,0}$  and $\rho_{c,0}$ [recall Eq.~\eqref{eq:N_scales}]. Plugging the above definitions in Eq.~\eqref{eq:gravothermal_nondim} leads to the following relations between the temporal functions:
\begin{equation}\label{eq:X_c_relations}
	\tilde{M}_c=\tilde{r}_c^3\tilde{\rho}_c\;\;,\;\;\tilde{v}_c^2=\frac{\tilde{M}_c}{\tilde{r}_c}\;\;,\;\;\tilde{L}_c=\tilde{\rho}_c\tilde{r}_c\tilde{v}_c^{5-n}.
\end{equation}
We find it convenient to define a new time variable
\begin{equation}
	\tau_c(\tilde{t})=t_{c,0}\tilde{\tau}_c=t_{c,0}\frac{\tilde{M}_c\tilde{v}^2_c}{\tilde{L}_c}.
\end{equation}
The spatial equations then take the following form:
\begin{align}\label{eq:gravo_spat}
	&\frac{\partial M_*}{\partial r_*}=r_*^2\rho_*\;\;,\;\;\frac{\partial}{\partial r_*}\left(\rho_* v_*^2\right)=-\frac{M_*}{r_*^2}\rho_*\;\;,\;\;L_*=-r_*^2\rho_* v_*^{3-n}\frac{\partial v_*^2}{\partial r_*}\nonumber\\
	&\frac{\partial L_*}{\partial r_*}=r_*^2\rho_* v_*^2\left[c_1-c_2\frac{\partial \log\left({v_*^3}/{\rho_*}\right)}{\partial \log M_*}\right]\; ,
\end{align}
where $c_1$ and $c_2$ are separation-of-variable constants, analogous to the ones used in~\citetalias{Balberg:2002ue} and~\citetalias{LyndenBell:1980}, generalized to velocity-dependent interactions. These constants relate the spatial functions to the temporal ones and are required to satisfy
\begin{equation}\label{eq:def_c1_c2}
	c_1=-{\tau}_c\frac{\partial}{\partial {t}}\log\left(\frac{{v}_c^3}{{\rho}_c}\right)\;\;,\;\;c_2=-{\tau}_c\frac{\partial}{\partial {t}}\log {M}_c,
\end{equation}
written in terms of dimensionful quantities.

The above is in agreement with \citetalias{LyndenBell:1980} for the case of $n=4$ and with \citetalias{Balberg:2002ue} for $n=0$. 
In the left panels of Fig.~\ref{fig:c1c2_logslope_plots}, we show $c_{1,2}$ calculated using our numerical code. While $c_{1,2}$ do not change significantly, the central region of the halo enters the SMFP phase (denoted by dashed lines)
before $c_{1,2}$ reach a constant value for some values of $n$. $c_{1,2}$ do reach a constant value for the case of $n=3,3.7$. Importantly, the values of $c_{1,2}$ for $n=3.7$ agree with the self-similar numerical solution of~\citetalias{LyndenBell:1980}, represented by gray lines. This fact is reassuring, since the condition for the approximated form of $\tilde{\kappa}$ in Eq.~\eqref{eq:mn_eq} is most strictly violated (in the outer regions of the halo) for the case of large $n$. Also indicated by gray lines in the left panels of Fig.~\ref{fig:c1c2_logslope_plots} are the  predictions for $c_{1,2}$ for the case $n=0$, based on the self-similar solutions of~\citetalias{Balberg:2002ue}. Our numerical results are similar in magnitude~\citetalias{Balberg:2002ue}, though they do not reach a constant value for the case of $n\simeq0$.

\paragraph*{Predictions of the Self-Similar Solution:} 
\label{par:predictions_of_the_self_similar_solution}
We next derive predictions using the self-similar equations and compare them to our numerical findings.
The mathematical conditions for a self-similar solution to exist are discussed in~\citetalias{LyndenBell:1980} and do not alter when a velocity dependence is included. An important validity condition identified in~\citetalias{LyndenBell:1980} is that $\rho$ should be time independent infinitely far from the halo center, which leads to following requirement:
\begin{equation}\label{eq:rho_alpha}
	\frac{d\log\tilde{\rho}_c}{d\log \tilde{r}_c}=\lim_{r_*\to\infty} \frac{d\log\rho_*}{d\log r_*}=-\alpha,
\end{equation}
with $\alpha$ in the range $[2,2.5]$~\citepalias{LyndenBell:1980}. Since all the central values $\tilde{x}_c$ are normalized such that $\tilde{x}_c(0)=1$, Eq.~\eqref{eq:rho_alpha} requires $\tilde{\rho}_c=\tilde{r}_c^{-\alpha}$.
Using Eq.~\eqref{eq:X_c_relations}, the other central parameters are thus given by
\begin{equation}\label{eq:x_c_of_r_c}
	\tilde{\rho}_c=\tilde{r}_c^{-\alpha}\;\;,\;\;\tilde{M}_c=\tilde{r}_c^{3-\alpha}\;,\;\tilde{v}^2_c=\tilde{r}_c^{2-\alpha}\;,\;\tilde{\tau}_c=\tilde{r}_c^{2+(n-3)\frac{2-\alpha}{2}},
\end{equation}
demonstrating nicely that the velocity dependence, parametrized by $n$, explicitly affects only the scaling of the timescale (or luminosity). Using Eq.~\eqref{eq:x_c_of_r_c} in Eq.~\eqref{eq:def_c1_c2} gives the relation
\begin{equation}\label{eq:alpha_cs}
	\alpha=6\frac{c_2-c_1}{c_2-2c_1}.
\end{equation}
 
The scaling of $\tilde{\rho}_c$ and $\tilde{v}_c$ with $\tilde{r}_c$ given in Eq.~\eqref{eq:x_c_of_r_c} results with
\begin{equation}\label{eq:alpha_logslope_relation}
	\frac{d\log\tilde{\rho}_c}{d\log\tilde{v}_c}=\frac{2\alpha}{\alpha-2}.
\end{equation}
We can compare this expression with the slope of the lines in Fig.~\ref{fig:rhoc_vc_plots}.
The lines of different $n$ have small differences in slope, causing $\tilde{\rho}_c$, $\tilde{M}_c$, and $\tilde{v}_c$ in Eq.~\eqref{eq:x_c_of_r_c} to have a slight implicit dependence on $n$ through $\alpha$.
In the right panel of Fig.~\ref{fig:c1c2_logslope_plots}, we plot this log slope as a function of the central 1D velocity dispersion, normalized by $V_{\mathrm{max}}$. We see that indeed most lines reach a constant right after the core size reaches a maximum $(t=t_{\rm core})$. The thin gray lines highlight our agreement with the numerical results of \citetalias{LyndenBell:1980} and \citetalias{Balberg:2002ue} for $n=4$ and $n=0$, respectively. Additional discussion on our numerical results for $\alpha$ is provided in Appendix~\ref{sec:validity_of_the_lmfp_self_similar_solution}.

We identify the time of core collapse as the time when $\tilde{\rho}_c$ and $\tilde{v}_c$ diverge, while $\tilde{r}_c$ vanishes. From Eq.~\eqref{eq:x_c_of_r_c}, we see that $\tilde{\tau}_c$ also vanishes for $\tilde{r}_c \to 0$, given that the exponent of $\tilde{r}_c$ must be positive. Thus, we can define the collapse time $\tilde{t}_{\rm coll}$ through $\tilde{\tau}_c(\tilde{t}_{\rm coll})=0$. Using Eq.~\eqref{eq:x_c_of_r_c}, we express $\tilde{v}_c^3/\tilde{\rho}_c$ in Eq.~\eqref{eq:def_c1_c2} in terms of $\tilde{\tau}_c$, which leads to  a  simple equation for $\tilde{\tau}_c$, solved by $\tilde{\tau}_c=1-\tilde{t}/\tilde{t}_{\rm coll}$, where
\begin{equation}\label{eq:t_coll}
	\tilde{t}_{\rm coll}=\frac{1}{c_1}\frac{6-\alpha}{(3\alpha-2)-n(\alpha-2)}.
\end{equation}
The collapse times we obtain numerically, which are shown in Fig.~\ref{fig:rhoc_rc_vs_t_plots}, do not match precisely this prediction. This discrepancy is to be somewhat expected, given that both $c_1$ and $\alpha$ inferred from Fig.~\ref{fig:c1c2_logslope_plots} do not approach an asymptotic value before the core enters the SMFP regime. The values of $c_1$ and $\alpha$ calculated in \citetalias{LyndenBell:1980} and \citetalias{Balberg:2002ue} agree qualitatively with our results. Using their values in Eq.~\eqref{eq:t_coll} gives $\tilde{t}_{\rm coll}\simeq 431$ for $n=4$ and $\tilde{t}_{\rm coll}\simeq 439$ for $n=0$, both consistently larger than the collapse time we obtain numerically, $\tilde{t}_{\rm coll} \simeq 335$. Further discussion on the validity of the LMFP self-similar predictions is included in Appendix.~\ref{sec:validity_of_the_lmfp_self_similar_solution}.

Lastly, we use the self-similar equation to estimate the time when the core enters the SMFP regime. We define $\tilde{t}_{c,\rm LS}$ to be the time when $\kappa_{\rm LMFP}\simeq\kappa_{\rm SMFP}$ at the halo center. In dimensionless variables, this condition is equivalent to solving $\hat{\sigma}^{2} \tilde{\rho} \tilde{v}^{2} \tilde{K}_{3} \tilde{K}_{5}\simeq 1$ at the core [cf. Eq.~\eqref{eq:kappa_tild}]. Similar to the replacement $\tilde{K}_3\to\tilde{v}^n$, we make the replacement $\tilde{K}_3\tilde{K}_5\to\tilde{v}^{2\bar{n}}$, where
\begin{equation}\label{eq:barn_def}
	2\bar{n}=-\left.\frac{d\log\tilde{K}_3\tilde{K}_5}{d\log\tilde{v}}\right|_{\tilde{v}=1}.
\end{equation}
Together with Eq.~\eqref{eq:alpha_logslope_relation}, we can identify the long-to-short mean free path transition velocity $\tilde{v}_{c,\rm LS}$ via
\begin{equation}\label{delta_sigma}
	1\simeq\hat{\sigma}^{2} \tilde{\rho} \tilde{v}^{2} \tilde{K}_{3} \tilde{K}_{5}\simeq\hat{\sigma}^2(\tilde{v}_{c,\rm LS})^{2\delta}\;\;,\;\;\delta=1-\bar{n}+\frac{\alpha}{\alpha-2}.
\end{equation}
The transition thus happens approximately at $\tilde{v}_{c,\rm LS}=\hat{\sigma}^{-1/\delta}$. For $n=0$ and $n=3.7$, we have $\delta=12.5$ and $\delta=7.9$ respectively. Using the above equation with the parameters of Run \#1 in Table~\ref{table:run_params}, for example, we find $v_{\rm LS}\simeq0.85 V_{\rm max}$. Run \#1 is the $n=0$ line in the left panel of  Fig.~\ref{fig:rhoc_vc_plots}; the prediction $v_{\rm LS}\simeq0.85 V_{\rm max}$ is in excellent agreement with the point at which this line changes from solid to dashed (which represents the LMFP-to-SMFP transition). This predicted scaling is shown to agree quite well with all our numerical runs in the left panel of Fig.~\ref{fig:vls_sighat}

We also use the above result to show why entering the SMFP does not drastically change the collapse time. The scalings in Eq.~\eqref{eq:x_c_of_r_c} allow us to identify $\tilde{\tau}_{c ,\rm LS}$, the ``central'' time at long-to-short mean free path transition, and relate it to $v_{c,\rm LS}$:
\begin{equation}
    \tilde{\tau}_{c,\rm LS}=\left(\tilde{v}_{c,\rm LS}\right)^{-\beta}=\hat{\sigma}^{\beta/\delta}\;\;,\;\;\beta=3-n+\frac{4}{\alpha-2}.
\end{equation}
For $0<n<4$, we find $1.9<\beta/\delta<2.4$. Given that we choose parameters such that the initial NFW halo begins it evolution in the LMFP regime, we have $\hat{\sigma}\ll1$ and thus conclude $\tilde{\tau}_{c,\rm LS}\ll1$. Since $\tilde{t}/\tilde{t}_{\rm coll}=1-\tilde{\tau}_c$, we find that the short-to-long transition occurs roughly at $t_{\rm LS}\simeq t_{\rm coll}(1-\hat{\sigma}^{\beta/\delta})$, which is very close to $t_{\rm coll}$. In the right panel of Fig.~\ref{fig:vls_sighat}, we show explicitly how the core collapse time is slightly delayed when increasing $\hat{\sigma}$.

\section{Conclusions}
\label{sec:conclusions}
We have studied the gravothermal evolution of isolated SIDM halos using equations for spherically-symmetric hydrostatic equilibrium, mass conservation, and the first law of thermodynamics, supplemented with a well-explored ansatz for heat transfer. We allow for velocity dependence in the cross section, which is natural in concrete particle physics models for SIDM. Our key finding is that these equations admit an approximate universality that is insensitive to the underlying particle physics model in the long mean free path regime. The universality allows a velocity-dependent cross section model to be systematically mapped onto a constant cross section model. 

We find that an NFW halo evolves to a maximal core size $r_{\rm core,0}\simeq0.45r_s$ when the core density and 1D velocity dispersion are given by  $\rho_{c,0}\simeq2.4\rho_s$, $v_{c,0}\simeq0.64V_{\rm max}$, respectively. In all cases, we find that the halo eventually enters the core collapse phase and transitions from the LMFP to the SMFP regime. To a good approximation, we find that the cross section only effects the timescale for gravothermal evolution and the time the central region of the halo enters the SMFP regime, both of which we explicitly quantify. The temporal evolution is dominated by the LMFP regime in large parts of model space. For these models, we have identified the dependence of the core collapse timescale on the halo properties and particle physics model in Eq.~\eqref{eq: t_c0_units}. We have found that the core collapse timescale depends on an average cross section defined at $v_{c,0}$ and is only very weakly dependent on the velocity dependence of the cross section (less than 3\% variation).

We note that the velocity average that sets the conductivity in the LMFP regime has no first principles derivation and needs to be determined from a comparison to N-body simulations. Our key results have been cast in terms of variables $\sigma_{c,0}$ and $n$, which can be easily computed for other averages that may turn out to be relevant in the LMFP regime. Similarly, $\sigma_{c,0}$ and $n$ could be trivially adapted to models other than the Yukawa interaction (under the Born approximation) we have adopted for this work. 

We have shown that self-similar solutions are possible in the LMFP regime for generic velocity-dependent cross sections and have provided a detailed derivation. Although these solutions are only approximately valid, they provide an intuitive explanation for many of the numerical results. They provide a good approximation to the core collapse timescale, but the variation in the timescale due to the velocity dependence (characterized by $n$) is larger in the self-similar solution than the numerical results. The self-similar solution also provides an approximate explanation of the structure and evolution of the inner density and dispersion profile of the halo. 

Our results imply that the thermal evolution of SIDM halos is nearly universal and vastly different cross section models can lead to the same SIDM halo in the LMFP regime. It also means that the analytic SIDM halo models developed for constant cross section cases can be easily adapted to the velocity-dependent cross section models. 

We have cast our result in terms of variables that are easily measured in simulations such as $V_{\rm max}$ and $r_s$. This should allow our results for the scattering (core-collapse) timescale and evolution of the central density and dispersion to be tested against N-body simulations of isolated and cosmological halos. It is natural to expect that there will be more scatter in the N-body simulations compared to the nearly universal solutions to the gravothermal equations. In such a case, we expect our equations, specifically the dependence on the cross section and halo parameters, to provide a good way to encapsulate the mean behavior of relevant densities, dispersions and time scales. Our results, when calibrated using N-body simulations, should provide a more robust method for using density measurements from galaxies and clusters to constrain the underlying scattering cross section. 

\section*{Acknowledgements}
\fix{We thank the anonymous referee for many useful and constructive comments and suggestions.} We thank Toby Opferkuch for many helpful comments on various aspects of this project. The work of NJO was supported in part by the Zuckerman STEM Leadership Program, the Azrieli Fellows Program and by the National Science Foundation (NSF) under the grant No.~PHY-1915314. K.~B.~acknowledges support from the NSF under Grant No.~PHY-2112884. M.~K.~acknowledges support from the NSF under Grant No.~PHY-1915005.

\vspace{-0.3cm}
\section*{Data Availability}
The processed data used for producing the figures in this paper are available upon request. The code used to produce the data will be released in a future publication.

\vspace{-0.3cm}

\bibliographystyle{mnras}
\bibliography{SMBH} 
\pagebreak


\appendix
{
\section{Velocity-dependent cross sections} 
\label{sec:over_counting}
In this appendix we discuss some of the complications that may arise when the SIDM interactions are velocity dependent. Specifically, we make a stronger argument for using $\sigma_{c,0}$ in semi-analytic models describing SIDM halo profiles. We first show that commonly used procedures lead to erroneous results when the interactions are velocity dependent. Based on this observation, we then make the case for mapping between velocity dependent and independent cross sections using $\sigma_{c,0}$ [c.f.Eq.~\eqref{eq:sig_c0}].
}

{
In the existing SIDM literature, it is commonly assumed that self-interactions impact the halo evolution only at radii smaller than $r_1$, defined as the radius below which SIDM particles interacted more than once during the age of the galaxy~\citep{Kaplinghat:2015aga}. The number of collisions an average particle at radius $r$ went through during the lifetime of a galaxy is roughly given by
\begin{equation}
	N_{\rm coll}(r)\sim\int_0^{t_{\rm age}}dt\rho(r,t)v(r,t)\frac{\sigma(v(r,t))}{m},
\end{equation}
where $\sigma$ is the \underline{total} cross section. $r_1$ is thus defined through $N_{\rm coll}(r_1)\sim 1$. While this simple definition is sufficient for the case of hard-sphere scattering, it is meaningless for the case of velocity dependent scattering. As an example, the Rutherford \underline{total} cross section diverges, driving $N_{\rm coll}\to\infty$ and, thereby $r_1\to 0$. A similar \fix{behaviour} occurs also for light but finite-mass mediators, \fix{where the number of collisions diverge as $1/m_{\rm med}^2$}. Thus, simply ``counting'' the number of collisions is not the right way to identify the region affected by self-interactions. 
}

{
The problem with the above approach is that thermalization requires efficient energy re-distribution, facilitated by collisional energy transfers. The Rutherford (or light Yukawa) cross section divergence is driven by forward collisions ($\theta\to0$), in which very little energy is being transferred. The energy transfer rate in Rutherford collisions is given by $\thb{n\sigma_T(v_{\rm rel})v_{\rm rel}E}$, where $E$ is the kinetic energy of the colliding particle and $\sigma_T=\int d\sigma(1-\cos\theta)$ is the momentum transfer cross section. While $\sigma_T$ also diverges, the divergence is only logarithmic and can be regulated in the far-forward direction with a small but nonzero minimum angle. The same kind of procedure is needed to regulate the Yukawa cross section. Note that $\sigma_T$ weights against far forward scattering ($\theta\to 0$), but not far backward scattering ($\theta\to\pi$).
Since the scattering of identical SIDM particles is symmetric under $\theta\to\pi-\theta$, we should consider the energy transfer rate in the collision of identical particles, $\thb{n\sigma_{\rm visc}(v_{\rm rel})v_{\rm rel}E}$~\citep{Sagunski_2021}, which also has a logarithmic divergence that must be regulated. Since $E\sim m_{\rm dm} v_{\rm rel}^2$, we take the energy transfer rate to scale as $\sim\thb{\sigma_{\rm visc}(v_{\rm rel})v_{\rm rel}^3}$. 
}

{
The above discussion suggests that $\thb{\sigma_{\rm visc}(v_{\rm rel})v_{\rm rel}^3}$ is the quantity governing the effect of self-interactions on halo evolution, and this expression is a function of the velocity dispersion.  The results presented in \S~\ref{sec:Core_Evolution} demonstrate that the LMFP evolution of halos is sensitive to particle physics only through $\sigma_{c,0}$. Up to an overall constant, $\sigma_{c,0}$ is precisely $\thb{\sigma_{\rm visc}(v_{\rm rel})v_{\rm rel}^3}$ evaluated at $v=0.64V_{\rm max}$. We conjecture that the velocity independent cross section results can be used for velocity dependent ones upon replacing the constant cross section with $\sigma_{c,0}$.
}

\section{Derivation of the Short Mean-Free-Path Heat Conduction} 
\label{sec:drivation_of_the_short_mean_free_path_heat_conduction}
In this appendix, we derive the heat conductivity term $\kappa$ in the SMFP regime, following mainly \S 10 of~\citep{pitaevskii2012physical}. We note that the notations in this appendix are slightly different than those used in the main text. Importantly, in this appendix $v=|\vecb{v}|$ is the magnitude of the velocity of a particle, \textbf{not} the 1D velocity dispersion. We begin with the non-relativistic Boltzmann equation
\begin{equation}\label{eq:boltzmann}
	\frac{df}{dt}=\left[\frac{\partial }{\partial t}+\boldsymbol{v}\cdot\boldsymbol{\nabla}+\boldsymbol{F}\cdot\frac{\partial}{\partial\boldsymbol{p}}\right]f=C[f].
\end{equation}
Here $C$ is the collision term, $\vecb{p}=m_{\rm dm}\vecb{v}$ and $f$ is the phase space density, normalized as such that
\begin{equation}
	\int d^3pf=n ,
\end{equation}
where $n$ is the number density.
The collision term for a 2-to-2 interaction is given by
\begin{equation}
	C[f](p_1)=\int v_{\rm rel}\frac{d \sigma}{d\Omega_{34}}\left[f(p_1)f(p_2)-f(p_3)f(p_4)\right]d^3p_2d\Omega_{34},
\end{equation}
where $v_{\rm rel}=|\boldsymbol{v}_1-\boldsymbol{v}_2|$ is the relative velocity between the incoming particles,  $p_{3,4}$ are the outgoing particles momenta, and $d\sigma/d\Omega_{34}$ is the differential cross section.
Momentum and energy conservation, given by $\boldsymbol{p}_1+\boldsymbol{p}_2=\boldsymbol{p}_3+\boldsymbol{p}_4$ and $E_1+E_2=E_3+E_4$, respectively, have been implicitly assumed. 
The structure of the collision term ensures that the system reaches a steady state ($df=0$) when $C[f]=0$, which is solved most generally by a Maxwell-Boltzmann distribution
\begin{equation}\label{eq:f_eq}
	f_{\rm eq}=\exp\left(\frac{\mu(T)-E}{T}\right)\;\;,\;\;E=\frac{m_{\rm dm}}{2}(\boldsymbol{v}-\boldsymbol{V})^2,
\end{equation}
where $T$ is the DM fluid temperature, $\mu$ is the chemical potential and  $\boldsymbol{V}=\thb{\boldsymbol{v}}$ is the bulk motion of the fluid.

\subsection{A small departure from equilibrium} 
\label{sub:a_small_departure_from_equilibrium}
We assume a small temperature gradient exists in the fluid, which results in a small departure from equilibrium. We, therefore, parametrize $f$ as follows
\begin{equation}
	f=f_{\rm eq}\left(1+\frac{\chi}{T}\right),
\end{equation}
where $\chi\ll T$ represents a small deviation from the equilibrium distribution $f_{\rm eq}$, for which the collision term vanishes identically $C[f_{\rm eq}]=0$. In what follows, we seek a perturbative expansion in $\chi/T$ that allows for the calculation of the heat conductivity. We assume that locally macroscopic quantities are given by the local equilibrium density, meaning
\begin{equation}
	\int d^3pf(p)\begin{pmatrix}1\\\boldsymbol{p}\\E(p)\end{pmatrix}=\int d^3pf_{\rm eq}(p)\begin{pmatrix}1\\\boldsymbol{p}\\E(p)\end{pmatrix},
\end{equation}
which leads to the following constraints on $\chi$:
\begin{equation}\label{eq:conservation_chi}
	\int d^3pf_{\rm eq}(p)\chi(p)\begin{pmatrix}1\\\boldsymbol{p}\\E(p)\end{pmatrix}=0.
\end{equation}
It can be shown (e.g., see~\cite{pitaevskii2012physical}) that to leading order in $\chi/T$, the Boltzmann equation takes the form
\begin{align}\label{eq:linear_thermo_coll}
	\left(\frac{\partial}{\partial t}+\boldsymbol{v}\cdot\boldsymbol{\nabla}\right)\chi&\equiv I[\chi]=\frac{E-c_PT}{T}\boldsymbol{v}\cdot\boldsymbol{\nabla}T+\nonumber\\
	&+\frac{1}{2}\left(m_{\rm dm} v_iv_j-\frac{E}{c_V}\delta_{ij}\right)\left(\frac{\partial V_i}{\partial x_j}+\frac{\partial V_j}{\partial x_i}\right),
\end{align}
where $c_P$ and $c_V$ are the specific heat capacity at constant pressure and volume, respectively, and $I$ is the linearized collision term
\begin{equation}\label{eq:I_chi}
	I[\chi](p)=\int v_{\rm rel}\frac{d \sigma}{d\Omega_{34}}\left[\chi(p)+\chi(p_2)-\chi(p_3)-\chi	(p_4)\right]d^3p_2d\Omega_{34}.
\end{equation}

\subsection{Heat flow and conduction} 
\label{sub:heat_flow_and_conduction}
The energy flux in a fluid is obtained by multiplying Eq.~\eqref{eq:boltzmann} by $E(p)=p^2/2m_{\rm dm}$ an integrating over $d^3p$, resulting in
\begin{equation}
	\frac{\partial (n {\cal E})}{\partial t}+\boldsymbol{\nabla}\cdot\boldsymbol{q}=0,
\end{equation}
where the mean energy $\cal E$ and the heat flux vectors $\vecb{q}$ are given by
\begin{equation}\label{eq:def_heat_flux}
	n{\cal E}=\int d^3p f(p)E(p)\;\;,\;\;\boldsymbol{q}=\int d^3p f(p)\boldsymbol{v}E(p).
\end{equation}
The heat conduction quantifies the response of the density function to small  temperature gradients. We, therefore, neglect shear terms (second line) in  Eq.~\eqref{eq:linear_thermo_coll} and arrive at the equation
\begin{equation}
	I[\chi]=\frac{E-c_PT}{T}\boldsymbol{v}\cdot\boldsymbol{\nabla}T.
\end{equation}
Since the above holds true for any (sufficiently small) $\boldsymbol{\nabla}T$, the linearity of $I$ [Eq.~\eqref{eq:I_chi}] motivates the  parametrization $\chi=G\boldsymbol{v}\cdot\boldsymbol{\nabla}T$, with which the equation above takes the form 
\begin{equation}\label{eq:BEQ_Gv}
	I[G\boldsymbol{v}]=\left(\frac{E}{T}-\frac{5}{2}\right) \boldsymbol{v},
\end{equation}
where we have used the heat capacity of an ideal gas. Using Eq.~\eqref{eq:def_heat_flux} and the definition of $\chi$ and $G$, the heat flux at leading order is
\begin{equation}
	q_{j}=\frac{\partial T}{\partial x_{i}} \frac{1}{T} \int d^{3} p f_{\mathrm{eq}} G(p) E(p) v_{i} v_{j} \equiv-\kappa_{i j} \frac{\partial T}{\partial x_{i}},
\end{equation}
where we have also used the constraints in Eq.~\eqref{eq:conservation_chi}. The heat conduction is defined in the momentary rest frame, where no preferred direction exist; therefore, we find
\begin{equation}\label{eq:kappa_th}
	\kappa_{ij}=\kappa \delta_{ij}\;\; ,\;\;\kappa=-\frac{m_{\rm dm}}{6T}\int d^3p f_{\rm eq}G(p)v^4=-\frac{\rho}{6T}\thb{G v^4},
\end{equation}
where $\thb{\cdot}$ denotes the thermal average with respect to the equilibrium distribution amd we have identified $m_{\rm dm}n=\rho$. The equation above allows us to calculate $\kappa$ once Eq.~\eqref{eq:BEQ_Gv} is solved.

\subsection{Approximated solution to the linearized Boltzmann Equation} 
\label{sub:approximated_solution_to_the_linearized_boltzmann_equation}
Following the spirit of the Chapman-Enskog expansion~\citep{pitaevskii2012physical,chapman1990mathematical}, we construct a complete set of polynomials $\left\{S_i(v^2)\right\}_{i=0}^{i=\infty}$, orthonormal with respect to thermal averaging in the following sense:
\begin{equation}
	\thb{S_i|S_j}\equiv\thb{v^{2}S_i(v^2)S_j(v^2)}=\delta_{ij},
\end{equation}
such that $S_i$ is a polynomial of order $i$. We next express both sides of Eq.~\eqref{eq:BEQ_Gv} as a series in the polynomials $S$
\begin{equation}\label{eq:S_expansion}
	G=\sum_{i=1}^\infty G_iS_i(v^2)\;\;,\;\;\frac{5}{2}-\frac{E}{T}= c_0S_0(v^2)+c_1S_1(v^2).
\end{equation}
$G$ doesn't contain an $S_0$ contribution to satisfy Eq.~\eqref{eq:conservation_chi}. The above definitions together with Eq.~\eqref{eq:kappa_th} allows us to write $\kappa$ as
\begin{equation}
	\kappa=-\frac{\rho}{6T}\thb{G v^4}=-\frac{\rho}{6T}\thb{G\left|v^2\right.}=-\frac{n}{3}\thb{G\left|\frac{E}{T}\right.}.
\end{equation}
Since $G$ does not include an $S_0$ contribution ($\thb{G|S_0}=0$), we may use Eq.~\eqref{eq:S_expansion} to replace $E/T\to E/T-5/2$. The orthonormality of our polynomials together with the expansion of $G$ thus gives
\begin{equation}
	\kappa=\frac{n}{3}\thb{G\left|\frac{5}{2}-\frac{E}{T}\right.}=\frac{n}{3}\thb{G|c_1S_1}=\frac{n}{3}c_1G_1.
\end{equation}
Our task is therefore reduced to finding $G_1$ and $c_1$.

Using a Gram–Schmidt procedure we can find $S_{0,1}$  
\begin{equation}
	S_0=\frac{1}{V_{\rm rms}}\;\;,\;\;S_1=\frac{1}{V_{\rm rms}}\sqrt{\frac{2}{5}}\left(\frac{5}{2}-\frac{E}{T}\right),
\end{equation}
Where $V_{\rm rms}=\thb{v^2}=3T/m_{\rm dm}$. The above allows us to easily identify
\begin{equation}
	c_0=0\;\;,\;\;c_1=\sqrt{\frac{5}{2}}V_{\rm rms}.
\end{equation}
Taking the dot product of Eq.~\eqref{eq:BEQ_Gv} with $\vecb{v}S_i$, thermal averaging, and using Eq.~\eqref{eq:I_chi} gives
\begin{equation}\label{eq:sum_rule}
	A_i=\thb{S_i\left|\frac{5}{2}-\frac{E}{T}\right.}=\frac{1}{n}\sum_j G_jI_{ij},
\end{equation}
with
\begin{align}
	I_{ij}&=-\int d^3pf_{\rm eq}S_i\boldsymbol{v}\cdot I[S_j\boldsymbol{v}]\nonumber\\
	&=\frac{1}{4}\int f_{\rm eq}(p)f_{\rm eq}(p_2)d^3pd^3p_2v_{\rm rel}\frac{d \sigma}{d \Omega}d\Omega \Delta[S_i\boldsymbol{v}]\Delta[S_j\boldsymbol{v}],
\end{align}
where in going from the first to second line, we use Eq.~9.8  from \S{}9 of chapter-I of~\cite{pitaevskii2012physical}, and $\Delta$ is defined as
\begin{equation}
	\Delta[F]=F(\boldsymbol{v}_3)+F(\boldsymbol{v}_4)-F(\boldsymbol{v})-F(\boldsymbol{v}_2).
\end{equation}
Note that by definition, $A_1=c_1=\sqrt{5/2}V_{\rm rms}$, while $A_{i\neq 1}=0$. To find $\kappa$ we still need to find $G_1$. This is now formally possible, but requires the inversion of an infinite dimension matrix $I_{ij}$. To leading order, the Chapman-Enskog expansion assumes that the sum in Eq.~\eqref{eq:sum_rule} can be truncated after the first term, {which has been shown to hold to within $\sim$3\% for the case oh hard spheres scatterings~\citep{chapman1990mathematical,Loyalka2007ChapmanEnskogST}, we discuss the validity of this truncation in the following subsection. To leading order we thus find}
\begin{equation}\label{eq:kappa_approx}
	nc_1=nA_1\simeq G_1 I_{11}\;\;\Rightarrow\;\;\kappa\simeq\frac{n^2 c_1^2}{3I_{11}}.
\end{equation}

Our calculation has reduced to calculating $I_{11}$. We recall that $c_1S_1(v_i^2)={5}/{2}-{E_i}/{T}	$, and note that momentum conservation gives $\Delta[\boldsymbol{v}]=0$. This allows us to find
\begin{equation}
	\Delta[\boldsymbol{v}S_1]=\frac{1}{c_1T}\Delta[\boldsymbol{v}E]=\frac{3}{2c_1 V_{\rm rms}^2}\Delta[v^2\boldsymbol{v}].
\end{equation}
It is convenient to calculate the above expression in the center-of-mass frame in which 
\begin{align}
	&\boldsymbol{v}=\boldsymbol{V}+\frac{1}{2}\boldsymbol{v}_{\rm rel}\;\;,\;\;\boldsymbol{v}_2=\boldsymbol{V}-\frac{1}{2}\boldsymbol{v}_{\rm rel}\nonumber\\
	&\boldsymbol{v}_3=\boldsymbol{V}+\frac{1}{2}\boldsymbol{v}_{\rm rel}'\;\;,\;\;\boldsymbol{v}_4=\boldsymbol{V}-\frac{1}{2}\boldsymbol{v}_{\rm rel}'.
\end{align}
Conservation of energy requires that $v_{\rm rel}=v_{\rm rel}'$. This leads to
\begin{equation}
	\Delta[v^2\boldsymbol{v}]=(\boldsymbol{v}_{\rm rel}\cdot \boldsymbol{V})\boldsymbol{v}_{\rm rel}-(\boldsymbol{v}_{\rm rel}'\cdot \boldsymbol{V})\boldsymbol{v}_{\rm rel}'.
\end{equation}
$I_{11}$ is only sensitive to  $\Delta[v^2\boldsymbol{v}]^2$. Moreover, the last term is the only one that depends on the direction of $\boldsymbol{V}$, and we average over it to find
\begin{equation}
	\int\frac{d \Omega_V}{4\pi}\Delta[v^2\boldsymbol{v}]^2=\frac{2}{3}V^2v_{\rm rel}^4\sin^2\theta,
\end{equation}
where $\theta$ is the scattering angle (angle between $\boldsymbol{v}_{\rm rel}$ and $\boldsymbol{v}_{\rm rel}'$) in the center-of-mass frame. Using the above in the definition of $I_{11}$, we find
\begin{equation}
	I_{11}=\frac{3n^2}{20V_{\rm rms}^6}\int d^3V V^2 f_V(V)\int d^3v_{\rm rel}f_{\rm rel}(v_{\rm rel})v_{\rm rel}^5 \sigma_{\rm visc}(v_{\rm rel}),
\end{equation}
where $f_V$ and $f_{\rm rel}$ are defined through
\begin{equation}
    n^2f_V(V) d^3Vf_{\rm rel}(v_{\rm rel}) d^3v_{\rm rel}=f_{\rm eq}(p_1) d^3p_1f_{\rm eq}(p_2) d^3p_2.
\end{equation}
The $V$ integral gives $V_{\rm rms}^2/2$, and we find
\begin{equation}
	I_{11}=\frac{3n^2}{40V_{\rm rms}^4}\thb{v_{\rm rel}^5 \sigma_{\rm visc}(v_{\rm rel})}_{\rm rel}.
\end{equation}
Plugging this back in Eq.~\eqref{eq:kappa_approx}, we obtain
\begin{equation}\label{eq:app_kappa_smfp}
	\kappa=\frac{100}{9}V_{\rm rms}^6\thb{\sigma_{\rm visc}(v_{\rm rel})v_{\rm rel}^5}_{\rm rel}^{-1},
\end{equation}
which concludes our derivation. Note that the averaging is carried over $v_{\rm rel}$ with
\begin{equation}
	f_{\rm rel}=\left(\frac{1}{4\pi v_{\rm 1D}^2}\right)^{3/2}\exp\left[-\frac{v_{\rm rel}^2}{4v_{\rm 1D}^2}\right]\;\;,\;\;v_{\rm 1D}^2=\frac{V_{\rm rms}^2}{3}=\frac{T}{m_{\rm dm}}.
\end{equation}

{
\subsection{Validity} 
\label{sub:validity}
Using the notation of the main text, Eq.~\eqref{eq:app_kappa_smfp} takes the form
\begin{equation}
	\kappa^{(1)}_{\rm SMFP}=\frac{3}{2}\frac{bv}{\sigma_0}\frac{1}{K_5}.
\end{equation}
As mentioned above, this expression holds true only to leading order in the Chapman-Enskog expansion, which we indicate explicitly with the superscript (1). If we were to expand to second order, similar steps to those presented above would lead to 
\begin{equation}
	\kappa^{(2)}_{\rm SMFP}=\frac{3}{2}\frac{bv}{\sigma_0}\frac{77K_5-112 K_7+80K_9}{28(K_5)^2+80K_5K_9-64(K_7)^2}.
\end{equation}
At the limit $w\to\infty$ all $K_n$ approach 1, and the above agrees with the results of~\citep{chapman1990mathematical,Loyalka2007ChapmanEnskogST}. We find that the ratio $\kappa^{(2)}_{\rm SMFP}/\kappa^{(1)}_{\rm SMFP}$ reaches a maximal value of $5/4$ at the $w\to0\;(n\to 4)$ limit. The minimal ratio obtained is $1$ at $w\simeq3.97v_{\rm 1D}$ and subsequently grow until it reaches $45/44$ at the hard spheres limit  $w\to\infty\;(n\to 0)$, in agreement with~\cite{Loyalka2007ChapmanEnskogST}. Including the third order led to an additional sub percent correction. We thus conclude that the leading order of the Chapman-Enskog expansion is valid in our case to within about 25\% accuracy. Since this paper is focused on the LMFP behavior of SIDM halos, we do not include second order corrections to our calculations.
}


\section{Short distance expansions of the LMFP solution} 
\label{sec:short_dist_expansion}
\begin{figure}
\centering
\includegraphics[width=0.48\textwidth]{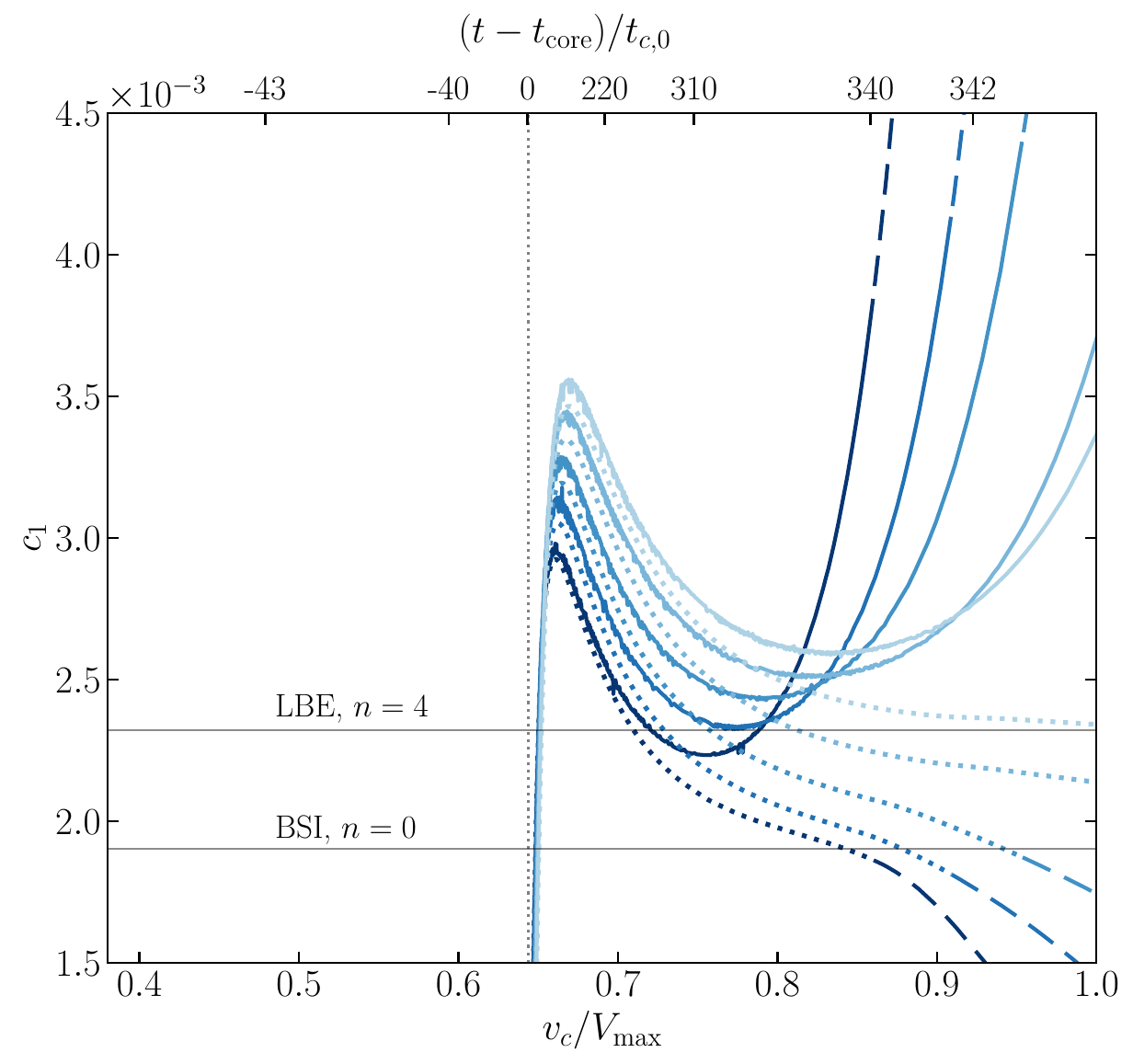}

\caption{
Separation of variables constant $c_1$ plotted as a function of the normalized central velocity dispersion. Dotted lines (Method 1) are calculated using Eq.~\eqref{eq:def_c1_c2} and are identical to the ones shown in Fig.~\ref{fig:c1c2_logslope_plots}. Solid lines (Method 2) are based on Eq.~\eqref{eq:c1_method2}, where we average over few of the smallest $r$ bins in our numerical results to minimize noise. The differences between the two methods demonstrate deviation from a self-similar  halo.
}
\label{fig:c1_comp}
\end{figure}

In this appendix, we derive the expansion of the self-similar solution in the limit $r_*\to 0$. Our starting point is Eq.~\eqref{eq:gravo_spat} with the initial conditions given in Eq.~\eqref{eq:BC_IC}. To leading order, we can most generally assume the following small-distance expansion:
\begin{equation}
    \rho_*=1+\delta_\rho r_*^{\beta_\rho}\;\;,\;\;v_*=1+\delta_vr_*^{\beta_v}\;\;,\;\;\beta_\rho,\;\beta_>0.
\end{equation}
The dimensionless mass satisfies $M_*'=r_*^2\rho_*$; therefore, to leading order, we have $M_*\simeq r_*^3/3$.
The luminosity $L_*$ satisfies
\begin{equation}
     L_*=-r_*^2\rho_* v_*^{3-n}\frac{\partial v_*^2}{\partial r_*},
\end{equation}
which gives
\begin{equation}
     L_*=-2\delta_v\beta_vr_*^{\beta_v+1}
\end{equation}
at leading order.
The {separation-of-variables constants} $c_1$ and $c_2$ are related to $L_*$ via the second line of Eq.~\eqref{eq:gravo_spat}:
\begin{equation}
        \frac{\partial L_*}{\partial r_*}=r_*^2\rho_* v_*^2\left[c_1-c_2\frac{\partial \log\left(\frac{v_*^3}{\rho_*}\right)}{\partial \log M_*}\right].
\end{equation}
To leading order, this gives
\begin{equation}
    \frac{dL_*}{dr_*}=-2\delta_v\beta_v(\beta_v+1)r_*^{\beta_v}=r_*^2c_1,
\end{equation}
where we drop terms of order ${\cal O}(r^{\beta_v+2},\;r^{\beta_\rho+2})$, since they are smaller than ${\cal O}(r^2)$.
Thus, the equation above gives
\begin{equation}
    \beta_v=2\;\;,\;\;c_1=-12\delta_v.
\end{equation}
The last term in Eq.~\eqref{eq:gravo_spat} is
\begin{equation}
    \frac{\partial}{\partial r_*}\left(\rho_* v_*^2\right)=-\frac{M_*}{r_*^2}\rho_*.
\end{equation}
To leading order this gives
\begin{equation}
    -\frac{r_*^2}{3}=\delta \rho \beta_\rho r_*^{\beta_\rho}+2\delta_v\beta_vr_*^{\beta_v}=\delta \rho \beta_\rho r_*^{\beta_\rho}+4\delta_vr_*^2,
\end{equation}
informing us that $\beta_\rho$ also equals $2$ and thus
\begin{equation}
    \rho_*\simeq1-\frac{1-c_1}{6}r_*^2\;\;,\;\;v_*\simeq1-\frac{c_1}{12}r_*^2\;\;,\;\;M_*\simeq\frac{r_*^3}{3}\;\;,\;\;L_*\simeq\frac{c_1}{3}r_*^3.
\end{equation}

\section{Validity of the LMFP self-similar solution} 
\label{sec:validity_of_the_lmfp_self_similar_solution}
This appendix is devoted to further analyze the the self-similar solution to the LMFP equations and to discuss its validity.

\begin{figure*}
\centering
\includegraphics[width=0.48\textwidth]{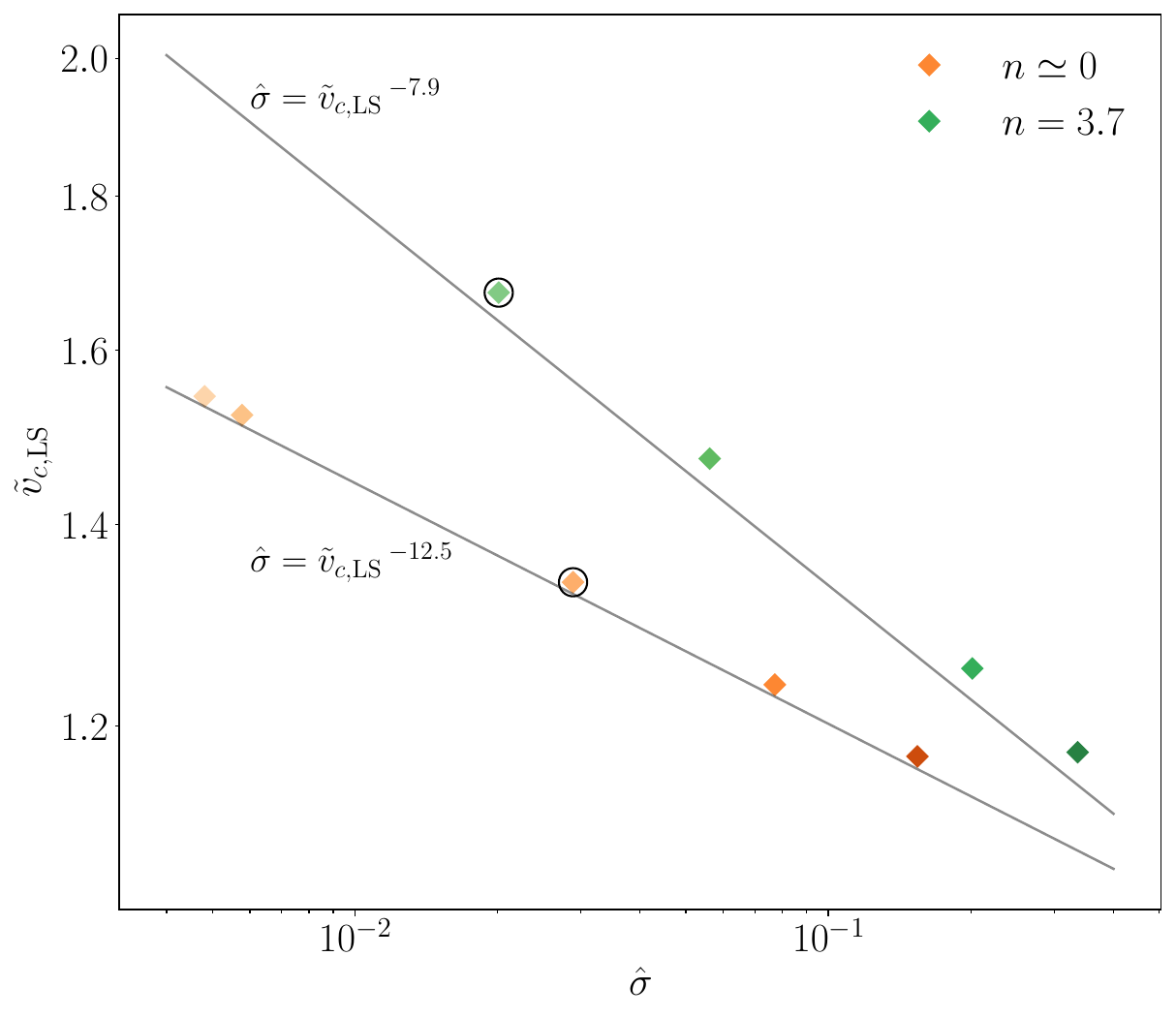}\;\;\;\;\includegraphics[width=0.5\textwidth]{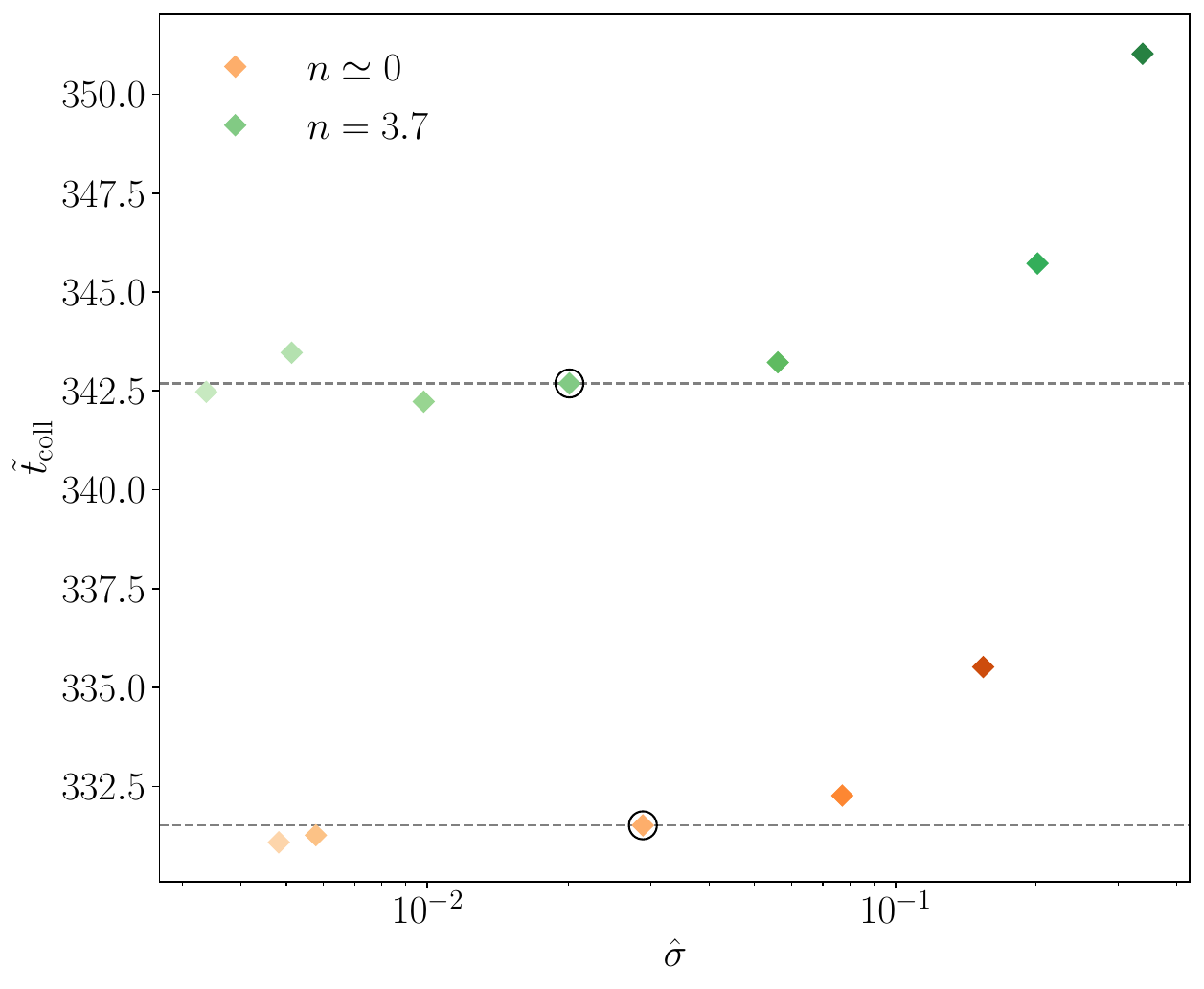}
\caption{ 
{\it (Left)} The central velocity at the time when ${\kappa_{\mathrm{SMFP}}}=\kappa_{\mathrm{LMFP}}$, normalized by $v_{c,0}$ and plotted  as a function of $\hat{\sigma}$ for different $n\simeq0$ and $n=3.7$ models. The gray lines show our predicted scaling, based on the self-similar solution, given by Eq.~\eqref{delta_sigma}. The numerical results agree very well with the predictions.
{\it (Right)} The collapse timescale, $\tilde{t}_{\mathrm{coll}}$, in our full numerical calculations, plotted as as a function of $\hat{\sigma}$ for $n\simeq0$ and $n=3.7$.  The dotted gray lines correspond to the $\tilde{t}_{\mathrm{coll}}$ shown in Fig.~\ref{fig:rhoc_rc_vs_t_plots}. The plot shows how core collapse is more delayed for increasingly larger values of $\hat{\sigma}$ (above a certain value). Circled diamonds indicate the models shown in the main text. See also Fig.~\ref{fig:rhoc_rc_vs_t_app} for the full time evolution of the same halos. {For the $n=3.7$ run, the points to the left of the circled diamond point are runs that did not reach SMFP, and as such, do not appear in the left panel.} 
}
\label{fig:vls_sighat}
\end{figure*}

\begin{figure*}
\centering
\includegraphics[width=0.98\textwidth]{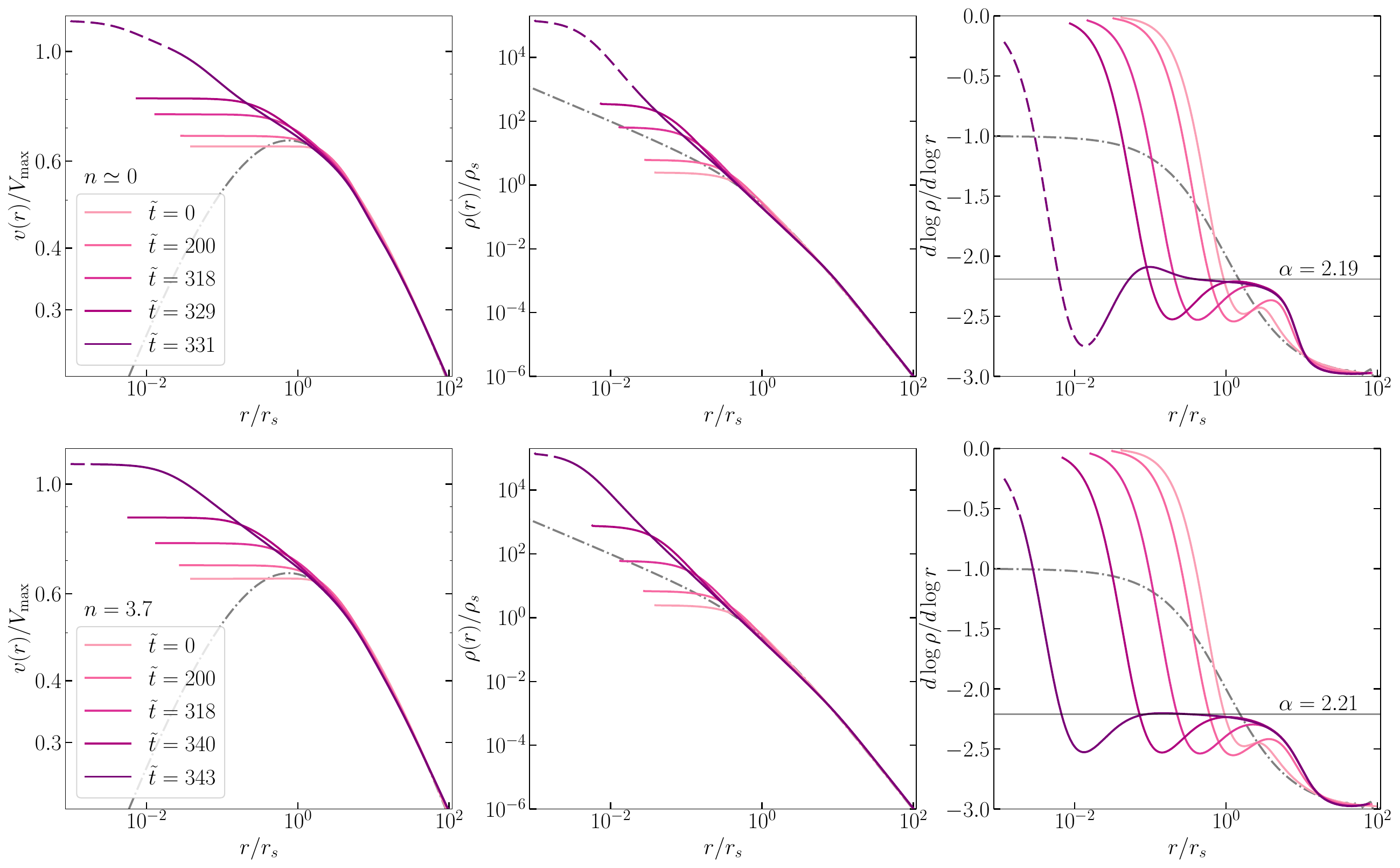}
\caption{
1D velocity dispersion profiles, mass density profiles, and log slopes of the mass density profiles for $n\simeq0$ {\it (top row)} and $n=3.7$ {\it (bottom row)}. The profiles are plotted for the time steps shown in the legend, with  $\tilde{t}=(t-t_{\rm core})/t_{c,0}$. The dashed regions of the profiles indicate where ${\kappa_{\mathrm{SMFP}}}<{\kappa_{\mathrm{LMFP}}}$, representing the parts of the halo evolving in the SMFP regime.  The dashed-dotted gray curves represent the initial NFW profiles. Solid gray lines in the third column of panels indicate values of $\alpha$ for $n\simeq0$ and $n=3.7$, derived using Eq.~\eqref{eq:alpha_logslope_relation} at time when $v_c=0.8V_{\rm max}$.}
\label{fig:v_rho_logslope}
\end{figure*}

In, the rightmost panels of Fig.~\ref{fig:v_rho_logslope}, we plot the log-slope of the density profile, $d\log \rho/d\log r$, at different times. Near $r\simeq r_s$, the halo eventually reaches a log-slope of $-\alpha$, indicated by the gray horizontal lines. The values of $\alpha$ given are obtained from the right panel of Fig.~\ref{fig:c1c2_logslope_plots}. This provides a direct confirmation of Eq.~\eqref{eq:alpha_logslope_relation}, which relates the temporal and spatial parts of the self-similar solution. Note that the profiles reach a slope of $-\alpha$ at times greater than $\tilde{t}\sim 300$, similar to the times at which the lines in Fig.~\ref{fig:c1c2_logslope_plots} reach a constant value. The self-similarity condition requires that the density is well-fit by the simple power law in Eq.~\eqref{eq:rho_alpha}, where $\alpha$ must be constant between $-2.5$ and $-2$~\citepalias{LyndenBell:1980,Balberg:2002ue}. However, the third column of Fig.~\ref{fig:v_rho_logslope} shows that at \textit{any} instance in the halo's evolution, the outer profile remains NFW-like with a slope $\sim-3$, which breaks the self-similarity condition. This is to be expected, since the interactions rate is suppressed at low densities, preventing the outer halo from thermalizing. We can also see that the self-similar solution is approximately achieved in the overall remarkable agreement with~\citetalias{Balberg:2002ue} and~\citetalias{LyndenBell:1980} in the right panel of Fig.~\ref{fig:c1c2_logslope_plots}.

The fact that the solution is approximate is also evident in the numerical values of the separation constants $c_{1,2}$ as plotted in the left panel of Fig.~\ref{fig:c1c2_logslope_plots}, which reach an approximate constant value only for the case of $n=3.7$ and $n=3$. Further, in Appendix~\ref{sec:short_dist_expansion}, we find
\begin{equation}
    \rho_*\simeq1-\frac{1-c_1}{6}r_*^2\;\;,\;\;v_*\simeq1-\frac{c_1}{12}r_*^2,
\end{equation}
which can be used to show
\begin{equation}\label{eq:c1_method2}
    \lim_{r\to 0}\left(\frac{d\log v^2}{d\log \rho}\right)_{t}=\lim_{r\to 0}\frac{(\partial\log v^2/\partial r)_t}{(\partial\log \rho/\partial r)_t}=\frac{c_1}{1-c_1}\simeq c_1.
\end{equation}
The subscript $t$ denotes a derivative at fixed time. This equation relates the temporal evolution of the core to the central gradients of velocity and density. The resulting $c_1$ is plotted together with the values of $c_1$, calculated using Eq.~\eqref{eq:def_c1_c2}. The two methods are compared In Fig.~\ref{fig:c1_comp}. While the methods agree up until around $0.75-0.80\,v_c/V_{\mathrm{max}}$, they diverge significantly afterwards and well before any systems reach the SMFP regime. These considerations imply that there is a limited range of time in the LMFP regime when the self-similar solution is approximately valid, but we do not have a controlled description of deviations from the self-similar solution in the LMFP regime.

Lastly, we wish to comment about the predicted collapse time of the halo core. Based on Eq.~\eqref{eq:t_coll}, the collapse time is predicted  to be $t_{\rm coll}\simeq 430t_{c,0}$, while numerically, we find $t_{\rm coll}\simeq 330$. Other than the fact that the prediction is correct only to within an order one factor, we find that the numerical results also differ in a qualitative manner from the predicted $t_{\rm coll}$. Eq.~\eqref{eq:t_coll} indicates that smaller values of $n$ correspond to later collapse times, but our numerical results suggest the opposite, as seen in Fig.~\ref{fig:rhoc_rc_vs_t_plots}. We suspect these discrepancies are due to the approximate nature of the self-similarity; however, it is not clear how deviations from self-similarity might effect our calculations.



\section{Parameters Table and additional figures}
\label{sec:figures_table}

\begin{landscape}
\begin{table*}
\centering
\begin{tabular}{||c c c c c c c c c c c c c c c||} 
 \hline
Run & $\rho_{s}$ & $r_{s}$ & $V_{\textrm{max}}$ & $\frac{\sigma_{0}}{m_\textrm{dm}}$ & $w$ & $\rho_{c,0}$ & $v_{c,0}$ & $\frac{\rho_{c,0}}{\rho_s}$ & $\frac{v_{c,0}}{V_{\mathrm{max}}}$ & $\frac{\sigma_{c}}{m_\textrm{dm}}$ & $\frac{\sigma_{100}}{m_\textrm{dm}}$ & $n$ & $\hat{\sigma}$ & $\hat{w}$ \\ 

  & $\left[10^7 \frac{M_\odot}{\textrm{kpc}^{3}}\right]$ & $[\textrm{kpc}]$ & $[\frac{\textrm{km}}{\textrm{s}}]$ & $[\frac{\textrm{cm}^2}{\textrm{g}}]$ & $[\frac{\textrm{km}}{\textrm{s}}]$ & $[10^7 \frac{M_\odot}{\textrm{kpc}^3}]$ & $[\frac{\textrm{km}}{\textrm{s}}]$ & &  & $[\frac{\textrm{cm}^2}{\textrm{g}}]$ & $[\frac{\textrm{cm}^2}{\textrm{g}}]$ &  &  &  \\ [0.5ex]  
 
 \hline\hline

 1 & 2.0 & 3.0 & 45.9 & 5.0 & $10^4$ & 4.9 & 29.5 & 2.5 & 0.64 & 5.0 & 5.0 & $2.1\times10^{-4}$ & 0.029 & 339.5 \\
 \hline
 2 & 2.0 & 3.0 & 45.9 & $10$ & 95.1 & 4.9 & 29.5 & 2.5 & 0.6 & 4.9 & 0.64 & 1 & 0.025 & 3.2 \\
 \hline
 3 & 2.0 & 3.0 & 45.9 & $40$ & 40.2 & 4.9 & 29.5 & 2.5 & 0.64 & 5.3 & 0.24 & 2 & 0.023 & 1.4  \\
 \hline
 4 & 2.0 & 3.0 & 45.9 & $1\times10^3$ & 10.8 & 4.9 & 29.5 & 2.5 & 0.64 & 4.60 & 0.09 & 3 & 0.019 & 0.37 \\
 \hline
 5 & 2.0 & 3.0 & 45.9 & $4\times10^6$ & 1.0 & 4.9 & 29.51 & 2.45 & 0.64 & 5.6 & 0.06 & 3.7 & 0.020 & 0.034 \\
 \hline\hline
 6 & 4.0 & 4.0 & 86.5 & 5 & $10^4$ & 9.8 & 55.5 & 2.45 & 0.64 & 5.0 & 5.0 & $7.4\times10^{-6}$ & 0.077 & 180.1 \\ 
 \hline 
 7 & 2.0 & 3.0 & 45.9 & 1.0 & $10^4$ & 4.9 & 29.5 & 2.5 & 0.64 & 1.0 & 1.0 & $2.1\times10^{-4}$ & 0.0058 & 339.5\\ 
 \hline
 8 & 40 & 0.8 & 54.7 & 5 & $10^4$ & 98.1 & 35.1 & 2.5 & 0.64 & 5.0 & 5.0 & $3.0\times10^{-4}$ & 0.15 & 284.7 \\  
 \hline 
 9 & 40 & 0.8 & 54.7 & $4\times10^6$ & 1.0 & 98.0 & 35.2 & 2.5 & 0.64 & 5.6 & 0.06 & 3.7 & 0.056 & 0.028 \\
 \hline 
 10 & 2.0 & 3.0 & 45.9 & $4\times10^7$ & 1.0 & 4.9 & 29.5 & 2.5 & 0.64 & 55.4 & 0.60 & 3.7 & 0.20 & 0.034 \\ 
 \hline
 11 & 2.0 & 3.0 & 45.9 & $6.7\times10^7$ & 1.0 & 4.9 & 29.5 & 2.5 & 0.64 & 92.9 & 1.0 & 3.7 & 0.34 & 0.034 \\ 
 \hline
\end{tabular}
\parbox{18cm}{\caption{ 
Table of input parameters and output parameters, with Runs \#1-5 serving as the main runs in this paper. Our input parameters are $\rho_s,\,r_s,\,V_{\mathrm{max}}$, $\sigma_0/m_\textrm{dm}$, and $w$. Output parameters are $\rho_{c,0},\,v_{c,0}$, $\frac{\rho_{c,0}}{\rho_s},\,\frac{v_{c,0}}{V_{\mathrm{max}}}$, $\frac{\sigma_c}{m_\textrm{dm}}$, $\frac{\sigma_{100}}{m_\textrm{dm}}$, $n$, $\hat{\sigma}$, and $\hat{w}$. 
$\sigma_0$ and $w$ are defined in Eq.~\eqref{eq:dsig_dO}, and $\sigma_{c,0}$ is defined in Eq.~\eqref{eq:sig_c0}. A more physically  intuitive quantity is $\sigma_{100}$, which we define analogously  to $\sigma_{c,0}$: $\sigma_{100}=\sigma_0K_3(100\,{\rm km\,s^{-1}}/w)$.}}
\label{table:run_params}
\end{table*}
\end{landscape}

\begin{figure*}
\centering
\includegraphics[width=0.49\textwidth]{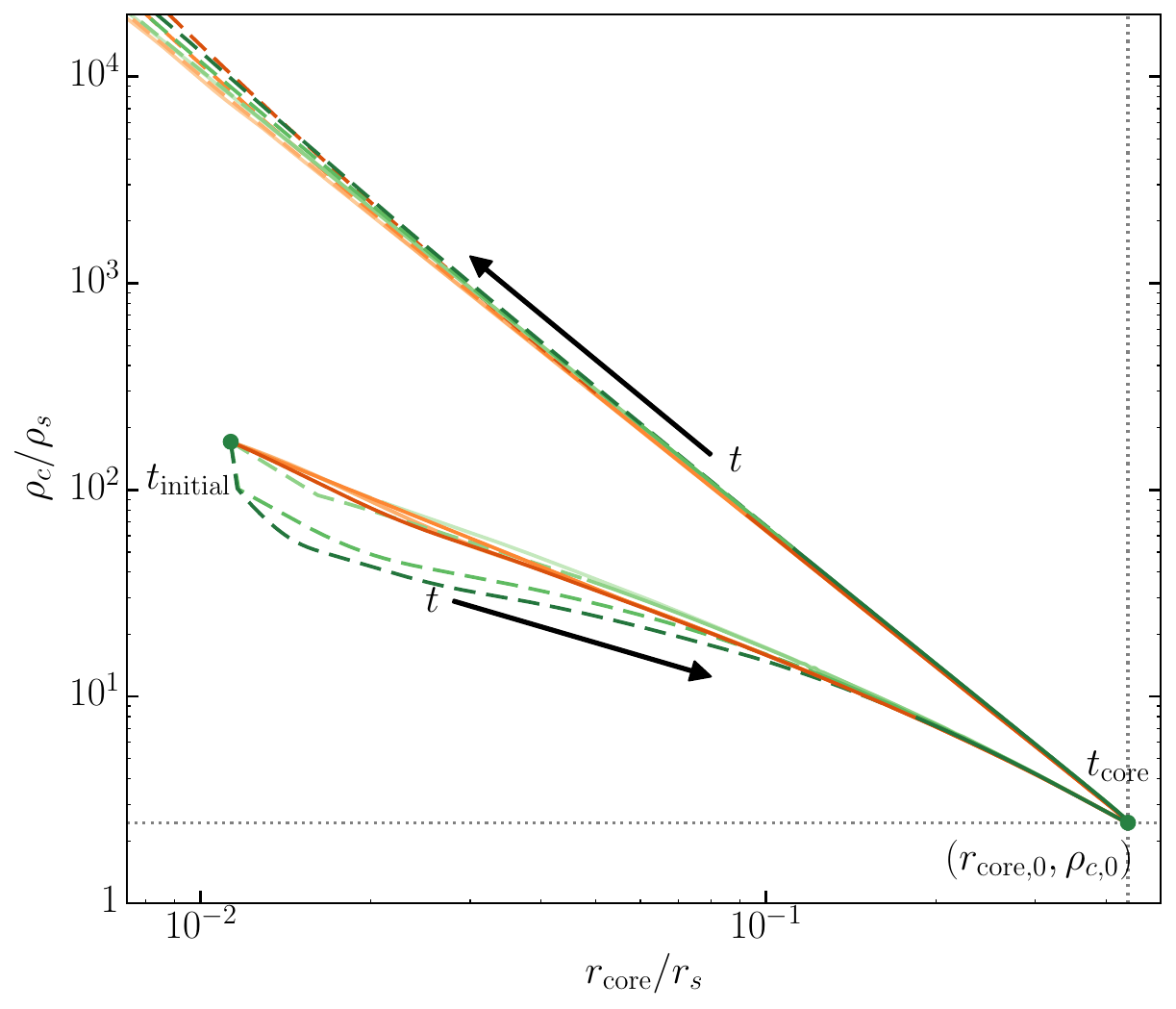}\;\;\;\;\includegraphics[width=0.49\textwidth]{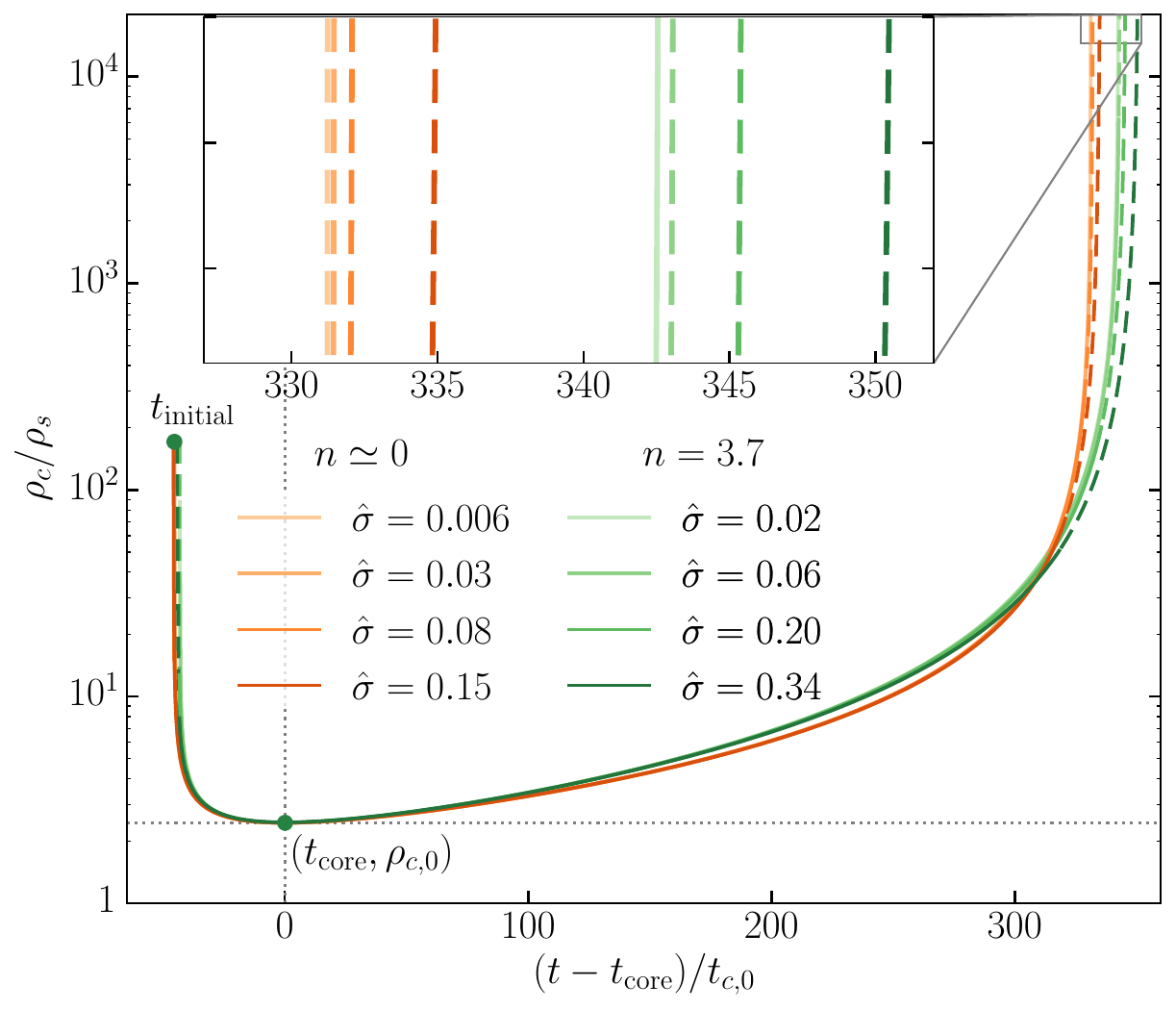}
\caption{
{\it (Left)} The normalized central density as a function of normalized core radius. 
{\it (Right)} Normalized central density as a function of normalized shifted time . All plotted quantities, labels and markers are the same as in the left panel of Fig.~\ref{fig:rhoc_rc_vs_t_plots}, except here we restrict our attention to the cases where $n\simeq 0$ and $n=3.7$ and highlight the dependence on $\hat{\sigma}$. Values of $\hat{\sigma}$ marked with a $*$ represent the ones shown in Fig.~\ref{fig:rhoc_rc_vs_t_plots}.  This figure does show a larger spread of the lines. Importantly, all lines reach the same maximal core with the same $r_{\rm core,0}$ and $\rho_{c,0}$. At later phases of the evolution, the lines diverge only after entering the SMFP, as indicated by dashed lines. The variation in collapse time due to late SMFP evolution is smaller than $3\%$.
}
\label{fig:rhoc_rc_vs_t_app}
\end{figure*}

\begin{figure*}
\centering
\includegraphics[width=0.49\textwidth]{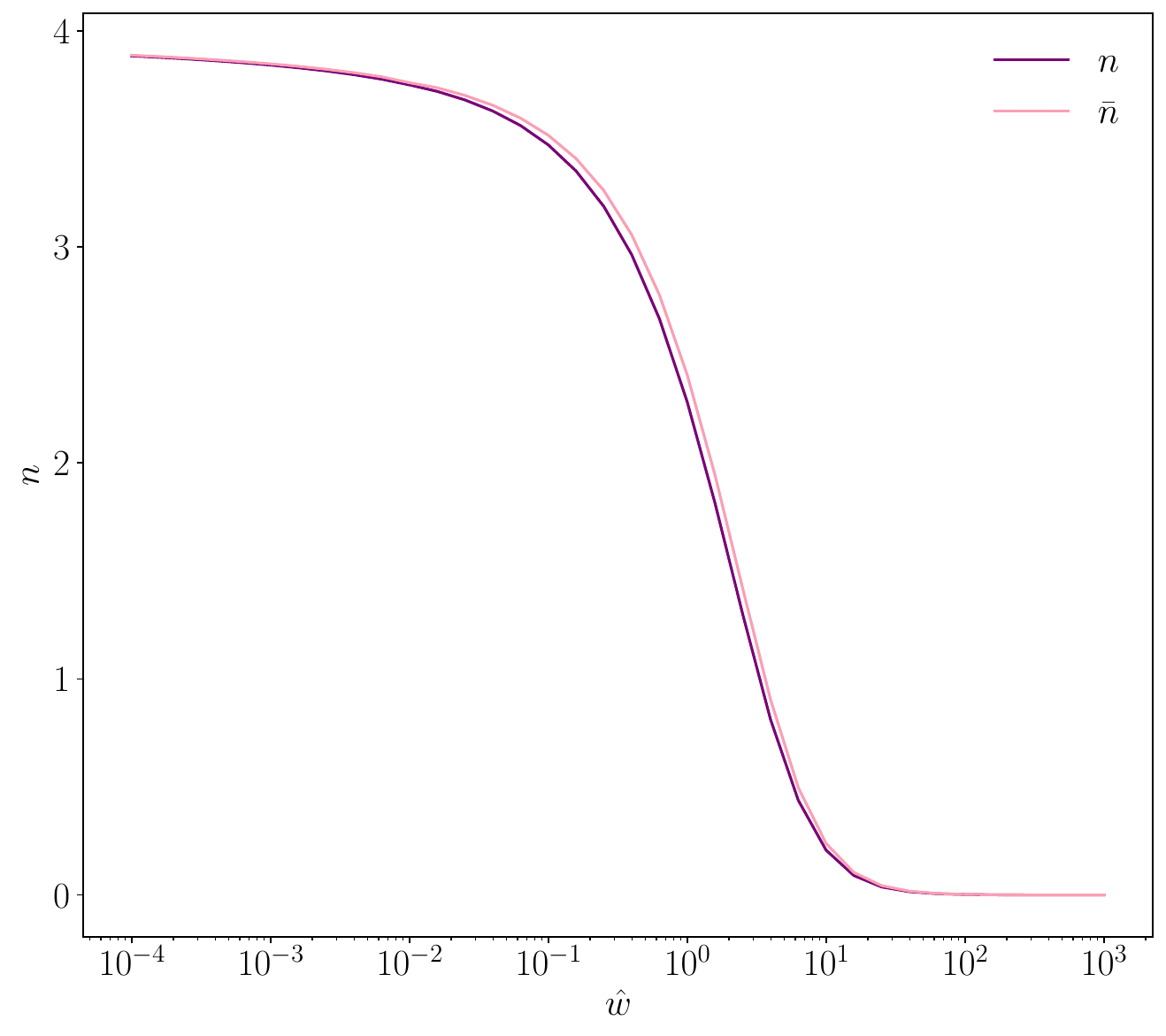}
\caption{The relation between $\hat{w}$, $n$ [Eq.~\eqref{eq:def_n}] and $\bar{n}$ [Eq.~\eqref{eq:barn_def}], for the Yukawa cross section in Eq.~\eqref{eq:dsig_dOmega}.
}
\label{fig:rhoc_rc_vs_t_app}
\end{figure*}



\end{document}